\documentclass[prb,twocolumn,floatfix,footinbib,superscriptaddress,showpacs,showkeys,notitlepage]{revtex4-1}
\usepackage{graphicx}
\usepackage[utf8]{inputenc}
\usepackage[english]{babel}
\usepackage{amsmath}
\usepackage{amssymb}
\usepackage{color}
\usepackage{hyperref}
\usepackage{caption}
\usepackage{subcaption}

\usepackage{array}
\usepackage{multirow}

\newcommand{\s}{\sigma}
\newcommand{\dd}{\dagger}
\newcommand{\up}{\uparrow}
\newcommand{\down}{\downarrow}

\newcommand{\melem}[3]{\ensuremath{\langle#1|#2|#3\rangle}}
\newcommand{\aver}[1]{\ensuremath{\langle#1\rangle}}
\renewcommand{\t}{\tau}
\newcommand{\w}{\omega}
\newcommand{\inttau}{\int_0^\beta d\t\ }
\newcommand{\intttau}{\iint_0^\beta d\t d\t'\ }

\DeclareMathOperator{\Tr}{Tr}

\begin{document}
\title{Slave rotor approach to dynamically screened Coulomb interactions in solids}
\author{I.~S.~Krivenko}
\affiliation{Centre de Physique Th\'{e}orique, Ecole Polytechnique, CNRS UMR7644, 91128
Palaiseau Cedex, France}
\affiliation{I. Institut f\"ur Theoretische Physik, Universit\"at Hamburg, Jungiusstraße 9,
20355 Hamburg, Germany}
\author{S.~Biermann}
\affiliation{Centre de Physique Th\'{e}orique, Ecole Polytechnique, CNRS UMR7644, 91128
Palaiseau Cedex, France}

\date{\today}

\begin{abstract}
Recent studies of dynamical screening of the electronic
Coulomb interactions in solids have revived interest
in lattice models of correlated fermions coupled to
bosonic degrees of freedom (Hubbard-Holstein-type models).
We propose a dynamical mean-field-based approach 
to dynamically screened Coulomb interactions.
In the effective Anderson-Holstein model, a transformation 
to slave rotors [S. Florens and A. Georges, Phys. Rev. B {\bf 66} 165111 (2002)] 
is performed to decouple
the dynamical part of the interaction.
This transformation allows 
for a systematic derivation and analysis of 
recently introduced approximate schemes for the
solution of dynamical impurity problems,
in particular, the Bose factor ansatz within
the dynamic atomic limit approximation (DALA) 
with and without Lang-Firsov correction.
More importantly still, it suggests an optimized
choice for a Bose factor
in the sense of the variational principle
of Feynman and Peierls.
We demonstrate the accuracy of our scheme
and present a comparison to calculations within
the DALA.
\end{abstract}


\maketitle

\section{Introduction}
\label{intro}

Lattice models of correlated fermions coupled to collective bosonic
modes have attracted renewed interest recently.
The Hubbard-Holstein
model\cite{HH2D} is a paradigmatic model to study the interplay 
between phonon-mediated attraction of
electrons and their electrostatic Coulomb repulsion\cite{ElPhReview}. 
The model exhibits a wide variety
of phenomena, such as bipolaron formation\cite{HHBipolaron1,HHBipolaron2},
charge density waves (CDW) and phonon-induced 
superconductivity\cite{HHCDWSC1,HHCDWSC2},
a metal-insulator (Mott) transition affected by the 
phonons\cite{HHMott1,HHMott2},
and non-Fermi-liquid behavior\cite{HHNonFL}.

Besides the more traditional context of electron-phonon coupling,
the Hubbard-Holstein model (or, equivalently, the Hubbard model with 
frequency-dependent interactions) has been acquiring new applications 
within realistic theories of correlated materials \cite{DALA,
werner-natphys2012}. 
Techniques based on density functional theory (DFT) are used to 
compute the one-particle
band structure of a material as a first step of combined numerical schemes such as
local-density approximation plus dynamical mean-field theory (LDA+DMFT) \cite{LDADMFT1,LDADMFT2} 
and LDA+$U$\cite{LDAU}. On the next step, electrons
from a subset of correlated bands are identified with fermionic degrees of freedom
of a lattice model. This downfolding procedure allows one to subsequently apply 
established many-body methods (in particular, DMFT)\cite{DMFT})
to account for correlation effects and obtain electronic
spectral functions \cite{EMSMAT, Tomczak_CompteRendus},
optical conductivities \cite{Tomczak2009_PRB,Tomczak2009_EPL}, 
or transport properties (see, e.g.,~\onlinecite{Tomczak-FeSi}).

This kind of calculation has recently acquired a new level of realism
thanks to techniques allowing for first-principles calculations
of the effective local Hubbard interactions also:
The constrained random-phase approximation (cRPA)\cite{cRPA1,cRPA2,cRPA3} 
even gives access to the energy-dependent matrix elements of the 
interaction. The energy dependence of this
Hubbard interaction $\mathcal{U}(\omega)$
reflects the fact that the high-energy itinerant states,
which are projected out from the full band structure, dynamically screen 
the Coulomb
interactions between correlated electrons. 
These screening processes
quite generally result in a substantial decrease of the static 
density-density interaction as parametrized by an energy-dependent
Slater parameter $F_0(\omega)$. They furthermore lead to renormalizations
of the one-particle hopping \cite{BandwidthRenorm}.
As a result, the phase diagram of the lattice model is substantially affected.

Dynamical impurity models are also a central ingredient of the
GW+DMFT method \cite{GWDMFT_BIERMANN_PRL2003,EDMFT_SUN}, 
combining many-body perturbation theory in the
framework of Hedin's GW approximation with DMFT.
Indeed, inspired by {\it extended} DMFT \cite{EDMFT}, this 
method maps a system with long-range interactions onto an
effective local problem with dynamical interactions, subject
to a double self-consistency condition relating one- and 
two-particle propagators to their counterparts in the solid.
The need for solving this dynamical effective impurity
problem has been a serious bottleneck hindering the
implementation of the scheme for quite some time, but
the recent development of the Bose factor ansatz (BFA) \cite{DALA}
as an efficient impurity solver
has finally unblocked the field \cite{tomczak_epl, tomczak_arxiv2013}.
For a review of the current status, see \cite{biermann_jpmc}.

Dynamical mean-field theory applied to the Hubbard-Holstein model
with local phonons
maps it onto an effective Anderson-Holstein model, parametrized by a 
bath hybridization
function $\Delta(i\w)$, which is subject to a self-consistency
condition.
Solving the effective impurity
model numerically amounts to obtaining its thermal Green's function,
$G(\tau) = - \langle \mathbb{T} d(\tau) d^{\dagger}(0) \rangle$, where
the average is performed by using the action corresponding to the
effective Anderson-Holstein model. Inclusion of bosons
into an impurity problem drastically enlarges the dimensionality 
of its state space, making it much more difficult to solve.  

In principle, it is still possible to apply continuous-time quantum Monte 
Carlo (CTQMC) algorithms to the Anderson-Holstein model.
In practice, and in particular for applications to realistic
materials, this is, however, not so straightforward:
The weak-coupling algorithm by Rubtsov\cite{CTINT} can treat 
interactions
with arbitrary frequency dependence, provided they are not too 
strong (in comparison to
the bandwidth) and do not contain high-frequency components. In general, 
neither of these
conditions is satisfied for realistic screened interactions -
effective impurity models are usually found in antiadiabatic or intermediate
regimes. A recently proposed generalization of the strong-coupling algorithm
by Werner and Millis\cite{CTHYB} does not suffer from these 
problems, and has allowed for several recent applications
within electronic structure calculations; see, e.g., 
\onlinecite{werner-natphys2012, Li_Huang_EPL2012_SrVO3}.
Nevertheless, there is still an infamous problem associated with the 
extraction of real-frequency
spectral data from imaginary-time results of a QMC run. 
A noticeable part of
the spectral weight (replica of the low-energy electronic structure) 
can lie far outside the bare electronic band,
when the impurity model is in the antiadiabatic regime. Resolution of 
spectra of this kind is notoriously difficult for analytic continuation tools.

Motivated by the mentioned problems, a new family of techniques has emerged
over the last few years.
The ``Bose factor ansatz'' (BFA) within the dynamic atomic limit approximation
(DALA)\cite{DALA} was proposed as an approximate - yet accurate - scheme capable
to circumvent the analytic continuation problem by reducing a given 
Anderson-Holstein problem to an effective Anderson model. The resulting Anderson model 
is readily solvable by existing impurity solvers, and the effect of the bosonic 
resonances is treated analytically within the BFA. 
Even the above-mentioned reduction of the effective bandwidth compared 
to its bare value
\cite{BandwidthRenorm}
can be taken into account: the change
can be estimated 
by means of a Lang-Firsov transformation\cite{LangFirsov1,LangFirsov2} and then 
incorporated into the DALA approach (DALA+LF) \cite{DALA}.

Following this route of study, in the present paper we introduce an
approach to the Anderson-Holstein model based on a slave rotor 
representation\cite{SlaveRotors}.
Slave rotor variables were introduced in \onlinecite{FlorensThesis,SlaveRotors}
as an efficient means to decouple charge and spin degrees
in low-energy models for correlated materials, even in the
case of dynamically screened interactions.
For impurity models in the DMFT context,
the formalism leads to a practical scheme allowing for an
approximate solution of the DMFT equations that correctly 
reproduces the Mott transition and the main spectral features
associated to it.
In contrast to previous work, however, we use the slave rotors 
here to decouple the dynamical part of the
density-density interaction only. It becomes clear within 
this framework, that DALA and DALA+LF can be understood as
simple approximations made on fluctuations of the slave phase field. 
Finally, we derive an effective mean-field modulation of
the hybridization function $\Delta(i\w)$ induced by the 
coupling to the phase variables, leading to an optimized
Bose factor ansatz in the sense of Feynman's variational
principle.

This paper is organized as follows. In Sec.~\ref{models}, we formulate the
Hubbard-Holstein model and an equivalent Hubbard model with energy-dependent
Hubbard interactions.
Within DMFT, this model is mapped onto an Anderson-Holstein model
with energy-dependent $\mathcal{U}(\omega)$.
In Sec.~\ref{rotors}, we analyze this model within a slave rotor 
transformation.
Different existing approximations (DALA, DALA+LF) are found to
be specific approximations to the slave rotors equations, as explained in Sec.~\ref{approximations}. 
A consequence of the coupling between lattice fermions and bosonic
degrees of freedom is a reduction of spectral weight in the low-energy
sector of the model, rationalized as an electronic polaron effect
\cite{BandwidthRenorm}. In Sec.~\ref{LF}, we
derive a finite-temperature generalization for the
expression of the bosonic renormalization factor $Z_B$
and compare it to an improved estimate within the slave rotor formalism.
Finally, a scheme beyond DALA and DALA+LF is proposed in Sec.~\ref{beyonddala}:
Derived from Feynman's variational principle, this 
optimized Bose factor ansatz contains both the high-energy plasmon
replica and the low-energy spectral weight reduction in a consistent
way. Section \ref{Bose-factor-DOS} contains a brief discussion 
of the physical meaning of the optimized Bose factor ansatz.
Section \ref{technical} summarizes the resulting self-consistency
loop and gives technical details.
In Sec.~\ref{results}, we present DMFT results for the Hubbard-Holstein model
obtained using our slave rotor scheme.
Finally, Sec.~\ref{conclusion} concludes the paper.
Two appendices present additional details concerning the derivation of
higher-order correlation functions within the slave rotor picture,
as well as of a temperature-dependent Lang-Firsov factor.

\section{Models with screened Coulomb interaction}
\label{models}

In the present paper, we will focus on 
the single-band Hubbard model with dynamically
screened Coulomb interactions $\mathcal{U}(\omega)$. 
Electrons can hop from site to site on a periodic lattice:
mathematically, electrons of spin $\sigma$
are created (annihilated) on site $i$
by operators $d^\dd_{i\s} (d_{i\s})$. The model is
defined by the finite-temperature action,
\begin{multline}\label{S_H}
    S_H = -\sum_{ij,\s} \inttau
	\bar d_{i\s}(\t)[(-\partial_\t +\mu)\delta_{ij} + t_{ij}] d_{j\s}(\t) +\\+
	U_\infty \sum_{i} \inttau n_{i\up}(\t) n_{i\down}(\t) +\\+
	\frac{1}{2}\sum_{i}\intttau N_i(\t) U_\textrm{ret}(\t-\t') N_i(\t').
\end{multline}

Here, $t_{ij}$ are hopping amplitudes between adjacent atoms on the lattice and 
$\mu$ is the chemical potential. The full density operators are defined as
$N_i = n_{i\up} + n_{i\down}$.
The instantaneous part of the interaction is denoted by $U_\infty$. Screening
is contained in the retarded part $U_\textrm{ret}(i\nu)$, which is chosen to be
negative and thus effectively 
reduces the on-site electron-electron repulsion.

The retarded interaction $U_\textrm{ret}(\t)$ can be represented as a
superposition of modes each parametrized by a position of a resonance
$\omega_\alpha$ and coupling strength $\lambda^2_\alpha$:
\begin{equation}\label{U_ret_tau}
    U_\textrm{ret}(\t) =
    -\sum_\alpha\lambda^2_\alpha
    \frac{\cosh[(\t-\beta/2)\omega_\alpha]}{\sinh[\omega_\alpha\beta/2]}.
\end{equation}
This expression is valid for $\t\in[0;\beta)$ and must be periodically 
continued outside the segment. We will also need a Matsubara frequency variant of this
expansion,
\begin{equation}\label{U_ret_inu}
    U_\textrm{ret}(i\nu) = 
    -\sum_\alpha \lambda^2_\alpha\frac{2\omega_\alpha}{\nu^2+\omega_\alpha^2}
\end{equation}
with bosonic Matsubara frequencies $\nu=\nu_n=n \frac{2 \pi}{\beta}$,
as well as an equivalent real-frequency description,
\begin{equation}
    U_\textrm{ret}(\t) =
    -\int\limits_0^{+\infty}\Im U_\textrm{ret}(\epsilon)
    \frac{\cosh[(\t-\beta/2)\epsilon]}{\sinh[\epsilon\beta/2]}
    \frac{d\epsilon}{\pi},
\end{equation}
which uses a screening spectral function $\Im U_\textrm{ret}(\epsilon) =
-\pi\lambda^2(\epsilon),\ \lambda(\epsilon) = \sum_\alpha \lambda_\alpha^2
[\delta(\epsilon-\omega_\alpha)-\delta(\epsilon+\omega_\alpha)]$.

The $\t$ dependence of the interaction makes it necessary to use the 
path-integral formalism and action (\ref{S_H}) instead of a Hamiltonian. However, 
in some cases,
it is more convenient to introduce a set of bosonic modes at each 
lattice site with
frequencies $\omega_\alpha$ and write a Hubbard-Holstein model Hamiltonian,
which is equivalent to $S_H$. In this case, the
frequencies $\omega_\alpha$ are
identified with plasmonic resonances (the ``charge cloud'' of the 
integrated out electrons plays the role of the plasma). 
The Hamiltonian includes a term which couples electrons to the introduced bosons,
\begin{multline}\label{H_HH}
    \hat H_\textrm{HH} =\\
	-\sum_{ij,\s} t_{ij} d^\dd_{i\s} d_{j\s}
	-\mu\sum_{i,\s} d^\dd_{i\s} d_{i\s}
	+ U_\infty \sum_{i} d^\dd_{i\up} d_{i\up} d^\dd_{i\down} d_{i\down}
	+\\+ \sum_{i,\alpha} \omega_\alpha b^\dd_{i\alpha} b_{i\alpha}
	+ \sum_{i,\alpha,\s} \lambda_\alpha
	    d^\dd_{i\s} d_{i\s} (b^\dd_{i\alpha} + b_{i\alpha}).
\end{multline}
The equivalence of $S_H$ and $\hat H_\textrm{HH}$ is readily 
verified by integrating out the $b^\dd_{i\alpha}, b_{i\alpha}$ variables.

It is worth noting that the Hubbard-Holstein model and the corresponding
action $S_H$ may be supplemented with additional site-local electron terms
(for instance, a local magnetic field), and most of the results of the present
paper will stand.

\section{Slave rotor transformation of the Anderson-Holstein model}
\label{rotors}

The slave rotor approach invented by Florens and Georges \cite{SlaveRotors} 
is an elegant and economic way to separate out and describe charge fluctuation 
in models of strongly correlated electrons. It was successfully applied to both impurity
\cite{FlorensThesis} and lattice models \cite{RotorsHubbard,RotortUJ},
to study the bandwidth-controlled and doping-controlled 
Mott transition \cite{RotorTransitions}, as well as to magnetism of multiorbital models
\cite{RotorMagnetism}.
Among other results, a slave rotor decoupling of the screened Coulomb interaction
$U(\t)$ was described by Florens in [\onlinecite{FlorensThesis}].
Here, we choose a different form of such a decoupling and provide a short reasoning
for the choice later in this section.

Within the slave rotor picture, one introduces a ``rotor phase'' variable 
$\theta$,
which is conjugate to the full charge, and a new pair of fermionic variables
$\bar f_\s, f_\s$ (called spinons hereafter). The rotor phase variable is 
related
to the Hubbard-Stratonovich scalar $\phi$ field, which is often used to 
decouple density-density interactions between electrons,
\begin{equation}\label{phi_theta}
    \phi(\t) \equiv \frac{\partial\theta}{\partial\t}, \quad
    \theta(\t) \in [0;2\pi), \quad
    \theta(0) = \theta(\beta),
\end{equation}
where the new fermion variables are proportional to the old ones with an
additional complex phase given by $\theta$,
\begin{equation}\label{d_f}
    f_\s \equiv d_\s e^{i\theta}, \quad \bar f_\s \equiv \bar d_\s e^{-i\theta}.
\end{equation}
The switch of variables $\phi\mapsto\theta$ and $\bar d,d\mapsto\bar f,f$ is
linear and thus the corresponding Jacobians of path integrals are irrelevant
constants. It is also worth noting that $\theta$ is introduced in such a 
way that there is no need to consider its static component: the resulting action and any
correlation function appearing in the theory contain either a $\tau$ derivative
of $\theta$ or a difference $\theta(\t)-\theta(\t')$.

The thermal Green's function of the original electrons is readily expressed in
terms of the new variables using definition (\ref{d_f}),
\begin{equation}\label{G_thru_Gf_GX}
    G(\t) \equiv -\aver{d_\s(\t) \bar d_\s(0)} =
    -\aver{f_\s(\t) \bar f_\s(0) e^{-i\theta(\t)+i\theta(0)}}.
\end{equation}

When the slave rotor transformation is used to study lattice models,
the degrees of freedom are usually introduced separately at each 
lattice site.
Some sort of a mean-field (saddle-point) approximation is then applied 
to the phase
variables $\theta_i$. This procedure allows for a decoupling of spinons and
chargons and to estimate the role of either of the two subsystems in a
given physical phenomenon.

Here, we use the slave rotor representation in a slightly different fashion.
Our approach is based on
dynamical mean-field
theory \cite{HHMott1, DMFT} (DMFT), mapping the lattice (Hubbard-Holstein)
model onto an 
effective Anderson
impurity model with frequency-dependent interactions $U(i\nu)$. 
This single-site
Anderson model is parametrized by a hybridization function $\Delta(i\omega)$,
which is to be determined self-consistently. The self-consistency condition is
dictated by the bare electronic band structure of the lattice
or crystal, i.e., by hopping matrix elements $t_{ij}$ (see Sec.
\ref{Bose-factor-DOS} for more details). The action of
the auxiliary Anderson model reads
\begin{equation}\label{S_AM}
    S_\textrm{AM} = S^\textrm{st}_\textrm{AM} + S^\textrm{dyn}_\textrm{AM},
\end{equation}
\begin{multline}\label{S_AM_st}
    S^\textrm{st}_\textrm{AM} = 
    -\sum_\s\inttau \bar d_\s(\t)[-\partial_\t + \tilde\mu] d_\s(\t) +\\+
    \sum_\s\intttau \bar d_\s(\t) \Delta(\t-\t') d_\s(\t') +\\+
    U_0 \inttau n_\up(\t) n_\down(\t),\\
\end{multline}
\begin{equation}\label{S_AM_dyn}
    S^\textrm{dyn}_\textrm{AM} = \frac{1}{2}\intttau N(\t) \bar U(\t-\t') N(\t').
\end{equation}

In the original action (\ref{S_H}) of the Hubbard model the full interaction function
is split into an unscreened part $U_\infty$ and a retarded part:
$U(\t) = U_\infty\delta(\t) + U_\mathrm{ret}(\t)$.
Here, in contrast, we have explicitly extracted the fully screened static component:
$U(\t) = U_0\delta(\t) + \bar U(\t)$, where
$U_0\equiv U(i\nu=0) = U_\infty - 2\sum_\alpha\lambda_\alpha^2/\omega_\alpha$
(and the chemical potential has
also undergone a modification, $\tilde\mu = \mu + \sum_\alpha \lambda_\alpha^2/\omega_\alpha$).
This has been done to proceed with a Hubbard-Stratonovich decoupling of $S^\textrm{dyn}_\textrm{AM}$ alone:
\begin{multline}\label{S_HS}
    S_\textrm{AM} =
    -\sum_\s\inttau \bar d_\s(\t)[-\partial_\t + \tilde\mu -i\phi(\t)] d_\s(\t) +\\+
    \sum_\s\intttau \bar d_\s(\t) \Delta(\t-\t') d_\s(\t') +\\+
    U_0 \inttau n_\up(\t) n_\down(\t) + S_\phi,
\end{multline}
\begin{equation}\label{S_phi}
    S_\phi = \frac{1}{2}\intttau \phi(\t) \bar U^{-1}(\t-\t') \phi(\t').
\end{equation}

Here, one sees the difference between 
the present approach and the approach by Florens \textit{et al}. 
In [\onlinecite{FlorensThesis}],
a $\phi$ field
was used to decouple the full interaction term in the Anderson model, including
both dynamical and static parts. In contrast to that, our intention is to 
associate
with rotors only the fluctuations caused by the dynamical part of the 
interaction.
As shown below, this choice of decoupling provides us with a more 
convenient description of dynamical screening. The physical effect of the
static component of the interaction is 
of a different nature than the finite-frequency components and
it is convenient to treat it separately.

An inverted operator $\bar U^{-1}(\t-\t')$ in the expression for $S_\phi$ should
not be understood in a literal mathematical way. Indeed, such an operator 
does not
exist, because the amplitude of the zeroth mode of $\bar U(\t-\t')$, i.e.,
$\bar U(i\nu=0)$, is zero by definition. To impart that operator a definite
meaning, we must include into the path integral only such trajectories
$\phi(\tau)$ that have no static component: 
$\inttau \phi(\t) = 0$ [$\phi(i\nu=0)=0$].

The substitution of spinon-chargon variables into (\ref{S_HS}) gives an 
expression for the
action which is explicitly split into three parts: an atomic part, 
a hybridization part, and the part with the dynamical interactions:
\begin{equation}\label{S}
    S =
    S_\mathrm{at}[\bar f,f] + 
    S_\mathrm{hyb}[\bar f, f;\theta] +
    S_\mathrm{dyn}[\theta],
\end{equation}
\begin{multline}\label{S_at}
    S_\mathrm{at}[\bar f,f] = -\sum_\s\inttau\bar f_\s(\t) [-\partial_\t
	+ \tilde\mu] f_\s(\t)
	+\\+ U_0 \inttau \bar f_\up(\t) f_\up(\t) \bar f_\down(\t) f_\down(\t),
\end{multline}
\begin{multline}\label{S_hyb}
    S_\mathrm{hyb}[\bar f,f;\theta] =\\=
	\sum_\s\intttau\bar f_\s(\t)\Delta(\t-\t')f_\s(\t')
	e^{i\theta(\t)-i\theta(\t')},
\end{multline}
\begin{equation}\label{S_dyn}
    S_\mathrm{dyn}[\theta] = \frac{1}{2}\intttau
	\partial_\t\theta(\t) \bar U^{-1}(\t-\t') \partial_{\t'}\theta(\t').
\end{equation}

This action plays a central role in the formalism being presented.
In the limit $\Delta(\t-\t')\to0$, fermionic
and rotonic degrees of freedom decouple, making the problem exactly solvable.
This is the strong-coupling limit of the theory, i.e.,
the dynamic atomic limit in terms of Ref. [\onlinecite{DALA}].
The thermal Green's function factorizes into fermionic
and rotonic parts if the corresponding degrees of freedom are not coupled in
the action:
\begin{eqnarray}
    G(\t-\t') = G_f(\t-\t') G_X(\t-\t'), \\
    G_f(\t-\t') \equiv -\aver{f_\s(\t) \bar f_\s(\t')}, \\
    G_X(\t-\t') \equiv +\aver{e^{i\theta(\t)}e^{-i\theta(\t')}}.
\end{eqnarray}

In this specific limit, the low-energy dynamics of fermions is 
determined by the screened interaction $U_0$, and
the particular form of $\bar U(\t-\t')$ enters only into a 
bosonic weight-modulating factor in the Green's function.
As we will discuss below, the dynamic atomic limit approximation
(DALA), as introduced in [\onlinecite{DALA}], corresponds to 
imposing a factorized form with the weight factor given by its
atomic limit expression even for finite $\Delta(\t-\t')$.
The slave rotor formalism is thus naturally suited for exploring
effects beyond the DALA.

\section{Approximations within the slave rotor picture}
\label{approximations}

In the atomic limit, $G_X$ can be calculated by a direct evaluation of the
corresponding path integrals,
\begin{equation}
    G^\textrm{at}_X(\t-\t') = \frac{\int\mathcal{D}[\theta] e^{i\theta(\t)-i\theta(\t')-S_\mathrm{dyn}[\theta]}}
	{\int\mathcal{D}[\theta] e^{-S_\mathrm{dyn}[\theta]}}.
\end{equation}
An expression for a general $2n$-point correlation function is derived in
Appendix \ref{corrfunctions}. Here we only give the result for the function 
of two times (see Fig. 1),
\begin{equation}\label{G_X_at}
    G^\textrm{at}_X(\t-\t') =
    \exp\left(-\frac{2}{\beta}\sum_{\nu>0}
    \frac{\bar U(i\nu)}{\nu^2}[1-\cos(\nu(\t-\t'))]\right).
\end{equation}
The argument of the exponential function in (\ref{G_X_at}) (denoted with $K(\t)$
in [\onlinecite{CTHYB}]) can be further
transformed by substituting Eq. (\ref{U_ret_tau}) and doing the Matsubara sum,
\begin{equation}\label{K_tau}
    K(\t) =
    \sum_\alpha \frac{\lambda_\alpha^2}{\w_\alpha^2}
    \frac{\cosh(\omega_\alpha(\t-\beta/2))-\cosh(\beta\omega_\alpha/2)}
        {\sinh(\beta\omega_\alpha/2)}.
\end{equation}

\begin{figure*}
    \includegraphics[scale=0.35]{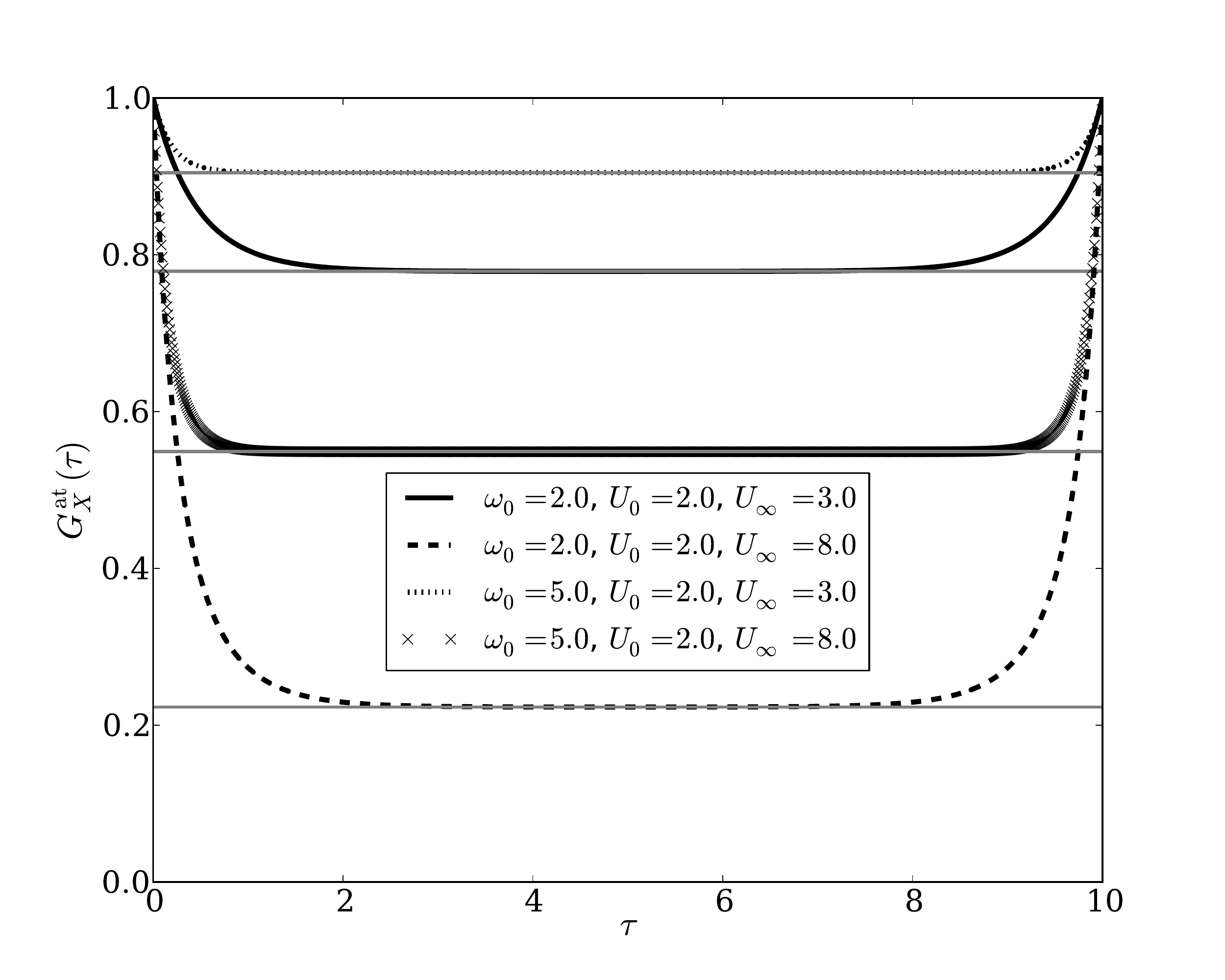}
    \includegraphics[scale=0.35]{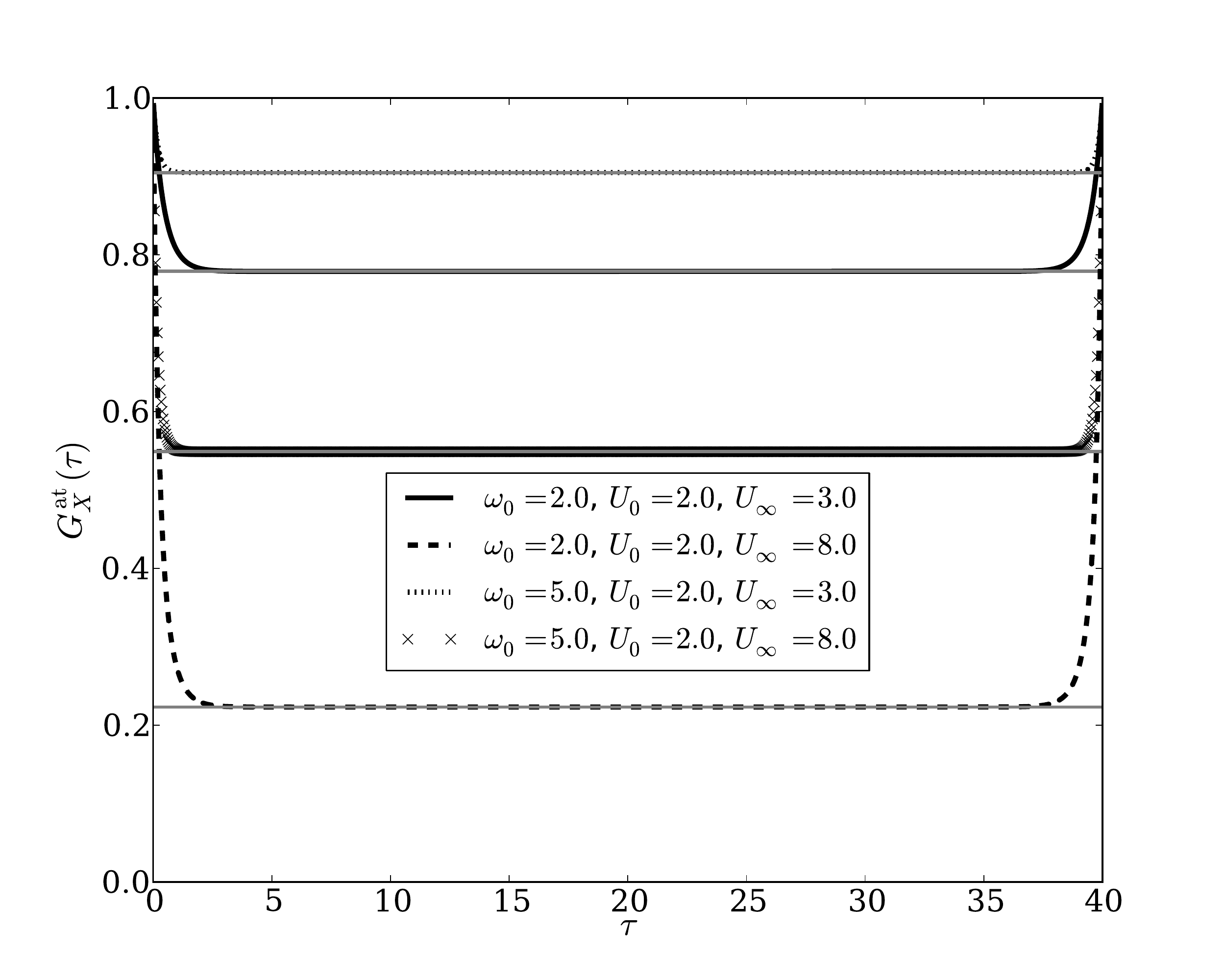}
    \caption{\label{G_X_figure}
    The rotor correlation functions $G_X^\textrm{at}(\tau)$ in the atomic limit 
    for the case of a single bosonic mode
    (left plot: $\beta=10$, right plot: $\beta=40$). Horizontal
    lines show the corresponding values of $Z_B$.}
\end{figure*}

This nontrivial exponential form of $G^\textrm{at}_X(\t)$ leads to an interesting
physical consequence. Let us consider a dynamical interaction function
$\bar U(i\nu)$ whose spectrum is localized around a single 
characteristic frequency
$\omega_0$. The spectrum of the auxiliary function $K(\t)$ will also have
this frequency as a special point. It is then readily seen from a Taylor
expansion of the exponent that all multiples of $\omega_0$ will be resonances
of $G^\textrm{at}_X(\t-\t')$; the spectrum of $K(\t)$ will be replicated along 
the 
frequency axis. For a more general case of several characteristic frequencies
$\omega_\alpha$, the spectrum of $G^\textrm{at}_X(\t)$ will exhibit features 
at all
combinatorial frequencies $\sum_\alpha m_\alpha \omega_\alpha$ with integer
coefficients $m_\alpha$.

Let us now switch on the hybridization of the impurity electrons with the
bath. $f$ fermions and
the rotor get coupled through $S_\mathrm{hyb}$ and the theory becomes nontrivial.
The simplest approximation to treat the model in this case is to 
artificially suppress
the coupling by putting $\theta(\t)-\theta(\t')$ to zero in $S_\mathrm{hyb}$
(the phase changes slowly over imaginary time).
This approximation was introduced in Ref. [\onlinecite{DALA}] as a means
to solve dynamic impurity models in the antiadiabatic limit at the cost
of static ones and was dubbed the ``dynamic atomic
limit approximation'' (DALA). 
In DALA, one neglects any mutual influence of $\Delta$
and fluctuations of the full charge, which are induced by the dynamical part of
the screened interaction. Obviously, this becomes a good approximation
when the energy scales separate, as is the case in the antiadiabatic
limit.
As follows from Eq. (\ref{G_thru_Gf_GX}), in DALA we have
\begin{equation}\label{G_DALA}
    G(\t)=G_f(\t;\Delta)G^\mathrm{at}_X(-\t),
\end{equation}
where $G^\mathrm{at}_X$ coincides with (\ref{G_X_at}) and $G_f(\t;\Delta)$ is
calculated for a conventional Anderson model with the given $\Delta$ and the
screened static interaction $U_0$. The factorized form (\ref{G_DALA})
corresponds to what is called the Bose factor ansatz (BFA) empirically 
introduced in [\onlinecite{DALA}] for the antiadiabatic limit. 
Interestingly, within the slave rotor
formalism, this factorization stands disregarding approximations made on
$\theta$ field and without limitations on the parameters of the 
Hubbard-Holstein model. 
This comes at the price, however, of introducing a coupling in the
differential equations governing the behavior of the two factors,
and, in this language, DALA simply corresponds to the zeroth-order
approximation in this coupling.

The advantage of DALA is that
it provides a simple way to reuse existing
quantum impurity solvers designed to work with purely 
static interactions. Moreover,
it substantially simplifies the ill-posed analytic continuation problem.
Extracting a spectral function $A(\epsilon)$ from noisy output data 
of a QMC run for the original
Anderson-Holstein model is a very difficult task. This is due to the presence
of the aforementioned plasmon satellites. On the other hand, the spectral 
function
$A_f(\epsilon)$ corresponding to $G_f(\t;\Delta)$ normally contains only 
low-energy scales [not larger than $\max$(bandwidth, $U_0$)]
and, for this reason, it is much easier to extract using a
maximum entropy algorithm. Thanks to Eq. (\ref{G_DALA}), the 
spectral function of physical electrons is a convolution,
\begin{equation}\label{convolution}
A(\epsilon) = \int_{-\infty}^{+\infty} d\epsilon'
    \frac{(1 + e^{-\beta\epsilon})A_f(\epsilon-\epsilon')B(\epsilon')}
    {(1+e^{\beta(\epsilon'-\epsilon)})(1-e^{-\beta\epsilon'})},
\end{equation}
where $B(\epsilon) = -(1/\pi)\Im G_X^\mathrm{at}(\epsilon)$ is the spectral function of rotors,
which may be calculated from (\ref{G_X_at}) at machine precision.

The main limitation of DALA is that it
overlooks the influence of high-energy bosonic fluctuations on low-energy 
dynamics of correlated spinons. 
For example, the coupling of the fermionic degrees of freedom to 
bosonic fluctuations leads to an ``electronic polaron effect,''
enhancing the mass of the effective low-energy fermionic degrees
of freedom or, equivalently, renormalizing the bare hopping matrix
elements. This effect has been investigated in detail in Ref. \onlinecite{BandwidthRenorm}
in the framework of Lang-Firsov transformation techniques.
As pointed out in Ref. \onlinecite{DALA},
the simplest way to refine DALA is to take into account an effective
an effective change in the magnitude of $\Delta(\tau)$ caused by the bosons.
Such a renormalization is described by the Lang-Firsov constant $Z_B$
(DALA+LF approximation).
As we will see below, the present framework lends
itself to an even more refined improvement: the slave rotor
formalism can be used to define a dynamical renormalization
of the hopping matrix elements, thus generalizing the simple
$Z_B$ renormalization.
Before explaining how to construct such a scheme, we will,
however, first need to introduce a generalization of the
simple $Z_B$ renormalization to finite temperatures.
This is done in the following section.

\section{Lang-Firsov transformation: effect of a finite temperature}
\label{LF}

When all bosonic resonances $\omega_\alpha$ lie far above energy levels of an
isolated atom and the boundary of the conduction
band $D$, the interplay of electronic and bosonic fluctuations may be accounted for
in a simplified way. This task may be accomplished through construction of a
low-energy effective model for electrons.

In this section, we present an effective model derivation based on the Lang-Firsov
transformation. This derivation is a finite-temperature generalization of the approach
proposed in Ref. [\onlinecite{BandwidthRenorm}].

The Lang-Firsov transformation is a unitary change of a basis in the state space
of $\hat H_\textrm{HH}$. It transforms the Hamiltonian and field operators,
replacing electrons $d^\dd_{i\s}, d_{i\s}$ with polarons $c^\dd_{i\s}, c_{i\s}$.
The unitary transformation operator is
\begin{equation}\label{LF_U}
    \hat U_\textrm{LF} = \exp\left(-\frac{\lambda}{\omega_0}\sum_{i,\alpha,\s} d^\dagger_{i\s}
    d_{i\s}(b_{i\alpha} - b^\dagger_{i\alpha})\right).
\end{equation}
The field operators transform as follows:
\begin{align}\label{LF_c}
    c_{i\s} &= \hat U_\textrm{LF} d_{i\s} \hat U_\textrm{LF}^\dd = 
    d_{i\s} \exp\left(
        \sum_\alpha\frac{\lambda_\alpha}{\omega_\alpha}
        (b_{i\alpha} - b^\dd_{i\alpha})
    \right)\\
    c^\dd_{i\s} &= \hat U_\textrm{LF} d^\dd_{i\s} \hat U_\textrm{LF}^\dd = 
    d^\dd_{i\s} \exp\left(
        -\sum_\alpha\frac{\lambda_\alpha}{\omega_\alpha}
        (b_{i\alpha} - b^\dd_{i\alpha})
    \right).
\end{align}
The Hamiltonian is completely equivalent to $\hat H_\textrm{HH}$, 
although it does not explicitly contain a coupling term between polarons and bosons
\cite{LangFirsov1,LangFirsov2}:
\begin{multline}\label{LF_H}
    \hat H_\textrm{LF} =
    \hat U_\textrm{LF} \hat H_\textrm{HH} \hat U_\textrm{LF}^\dd =\\=
    -\sum_{ij,\s} t_{ij} c^\dd_{i\s} c_{j\s}
    -\tilde\mu\sum_{i\s} c^\dd_{i\s} c_{i\s}
    +\\+ U_0 \sum_{i} c^\dd_{i\up} c_{i\up} c^\dd_{i\down} c_{i\down}
    + \sum_{i,\alpha} \omega_\alpha b^\dd_{i\alpha} b_{i\alpha}.
\end{multline}

The Lang-Firsov transformation results in a renormalization of the chemical
potential $\tilde\mu$ and the unscreened part of the interaction $U_0$;
these quantities coincide with those in (\ref{S_AM_st}).

The polaron degrees of freedom represent electrons dressed by bosonic fluctuations.
One can take this dressing into account in an approximate way by calculating
renormalized hopping constants $Z_B t_{ij}$ for the original electrons. In the paper
by Casula \textit{et al.} \cite{BandwidthRenorm} this 
is done by projecting the Lang-Firsov
Hamiltonian onto the subspace of zero-boson states,
\begin{equation}
    \hat H_\textrm{eff} = \melem{\{0\}_\alpha}{\hat H_\textrm{LF}}{\{0\}_\alpha},
    \textrm{ given }\omega_\alpha\gg U_0, t_{ij}.
\end{equation}

This leads to the following effective low-energy Hamiltonian:
\begin{multline}\label{H_eff}
    \hat H_\textrm{eff} = 
    	-\sum_{ij,\s} Z_B t_{ij} d^\dd_{i\s} d_{j\s}
	-\tilde\mu\sum_{i\s} d^\dd_{i\s} d_{i\s}
	+\\+ U_0 \sum_{i} d^\dd_{i\up} d_{i\up} d^\dd_{i\down} d_{i\down},
\end{multline}
which is a conventional Hubbard model with a renormalized bandwidth, interaction
strength, and chemical potential, but where the one-particle hopping matrix
elements have been renormalized by a factor $Z_B<1$ that we discuss below.

This simple renormalization is valid in the antiadiabatic regime and for not very
high temperatures. As the temperature goes higher, more boson excitations are
effectively created and the projection onto the zero-boson subspace becomes
no longer valid.

When the effective low-energy  model is used to calculate a spectrum of physical
electrons, the Green's function reads
\begin{equation}
    G^\textrm{low-energy}_{i\s;j\s}(\t) =
	-Z_B \aver{\mathbb{T} d_{i\s}(\t) d^\dag_{j\s}(0)}_{\hat H_\textrm{eff}}.
\end{equation}
The full spectral weight, corresponding to a Green's function defined in this
way, is equal to $Z_B$. The remaining 
contribution $1-Z_B$ 
to the weight is
constituted by high-frequency scattering processes involving boson
creations/annihilations. As we have excluded such processes from
$\hat H_\textrm{eff}$, consistency of the theory requires us to diminish the
spectral weight accordingly.

Here we present a finite-temperature expression for $Z_B$, which is obtained by
tracing out all bosonic degrees of freedom from $\hat H_\textrm{LF}$ (a complete
derivation is found in Appendix \ref{Z_B_finitet}),
\begin{multline}\label{Z_B}
    Z_B = \exp\left(-\sum_\alpha
        \frac{\lambda_\alpha^2}{\omega_\alpha^2}\coth(\beta\omega_\alpha/2)
    \right) =\\=
    \exp\left(
        \frac{1}{\pi}
            \int_0^{+\infty}
            \frac{\Im U_\textrm{ret}(\epsilon)}{\epsilon^2}
            \coth(\beta\epsilon/2) d\epsilon
    \right).
\end{multline}

Expression (\ref{Z_B}) contains two dimensionless combinations of three energy scales,
namely, $\lambda_\alpha/\omega_\alpha$ and $\beta\omega_\alpha$. They give rise to
a multitude of temperature limiting cases, some of which we analyze here.
\begin{itemize}
    \item $\beta\to\infty;\ \lambda,\omega=\textrm{const}$. This is the zero-temperature
        limit, consistent with the result of Casula \textit{et al.},
        $Z_B =\exp(-\sum_\alpha \lambda_\alpha^2/\omega_\alpha^2)$.
        
    \item $\beta\to0;\ \beta\omega=\textrm{const}$. If all energy scales of the
        bosonic subsystem follow a rise of the temperature, then our expression stands
        for arbitrarily small $\beta$. However, this case is usually of little physical
        interest.

    \item $\beta\to0;\ \lambda,\omega=\textrm{const}$. In this more physically relevant case,
        we get the curious result that $Z_B$ must vanish together with beta.
        The system rapidly falls down to the atomic limit as the temperature grows.
        It seems that $Z_B$ alone can provide a reasonable effective description
        only for not very small values of $\beta\omega_\alpha$. One can use
        the condition $Z_B\lesssim1$ to determine the order of magnitude of the temperature
        where a description in terms of $Z_B$ is still valid. Beyond this region,
        a more refined theory of dynamical screening is needed.

\end{itemize}

Of course, it is far from obvious that the calculated temperature
dependence of $Z_B$ plays any role in practical applications.
Indeed, for realistic plasmon frequencies
$\omega_\alpha\simeq 10\textrm{ eV}$, 
the temperature which could grant considerable
boson excitation probabilities would be too high to observe interesting
electron correlation effects. 
Nonetheless, the obtained expression will turn out to be useful for
finding a connection between the Lang-Firsov trick 
and the formalism presented below.

How can these insights be used within the slave rotor framework employed
as a solver technique for the DMFT equations with dynamical interactions?
A renormalization of the hopping matrix translates into a renormalization
of the hybridization function $\Delta(i\w)$, when a Hubbard-Holstein model
(or an equivalent Hubbard model with a dynamical interaction) is mapped onto the
Anderson model by DMFT. The effective renormalization of $\Delta(\t-\t')$ in the
slave rotors picture is caused by a coupling term,
$\Delta(\t-\t')e^{i\theta(\t)-i\theta(\t')}$. 
Obviously, the $Z_B$ factor thus corresponds to a specific approximation
to the second factor in this expression.
One could, for example, think of the following mean-field
estimate of the renormalization constant $Z_B$:
\begin{equation}
    Z_B = \frac{1}{\beta}\inttau \aver{e^{-i\theta(\t) + i\theta(0)}}_\textrm{at},
\end{equation}
where the subscript ``at'' indicates that the average value is taken
with the atomic limit action.

The imaginary-time integral in this expression means that we are interested only
in the low-energy fluctuations of the $\theta$ field. 
Substituting $G^\mathrm{at}_X(\t)$ into this expression, we rewrite the integral,
\begin{multline}\label{Z_B_rotors}
    Z_B = \exp\left(-\sum_\alpha
        \frac{\lambda_\alpha^2}{\omega_\alpha^2}\coth(\beta\omega_\alpha/2)
    \right) \times\\\times
    \int_{-1/2}^{+1/2} dx
    \exp\left[
        \sum_\alpha\frac{\lambda_\alpha^2}{\w_\alpha^2}
        \frac{\cosh(\beta\w_\alpha x)}{\sinh(\beta\w_\alpha/2)}
    \right].                     
\end{multline}
If the adiabatic ratios $\lambda_\alpha/\w_\alpha$ are all small, the right-hand
side integral goes to 1, and our estimate of $Z_B$ becomes consistent with
the antiadiabatic-limit expression (\ref{Z_B}).
While we could use the improved estimate (\ref{Z_B_rotors}) for $Z_B$ to try 
reaching the
$\lambda_\alpha/\w_\alpha\simeq 1$ region in practical calculations, we prefer to
go a bit further. In the next section, we introduce an approximation beyond DALA,
in which $\Delta$ undergoes a dynamic rather then a static renormalization.

\section{Beyond DALA: the optimal mean-field treatment of $S_\textrm{hyb}$}
\label{beyonddala}

We now come to the central idea of the present work. The above
discussion has addressed how dynamical screening leads to a mass
enhancement of the low-energy fermionic degrees of freedom,
as expressed by the bosonic renormalization factor $Z_B$.
It also became clear, however, that replacing the truly dynamical
couplings between fermionic and rotor degrees of freedom by a
simple number introduces simplifications that are justified only
in specific limits.
In this section, we will derive a more general {\it dynamical}
renormalization scheme, based on the slave rotor framework.
More precisely, we will construct an (in a sense to be specified)
{\it optimal} bosonic renormalization factor.

To this end, we consider a class of effective spinon-only models with
modified hybridization functions $\tilde\Delta(\t)$,
\begin{equation}
    \tilde S_\mathrm{hyb}[\bar f,f] = \intttau \sum_\s
    \bar f_\s(\t)\tilde\Delta(\t-\t')f_\s(\t').
\end{equation}

Whenever an action $S$ of a physical system is replaced by a simpler ``trial''
action $\tilde S$, it is convenient to apply Feynman's variational criterion to
estimate what is the best choice of the parameters of $\tilde S$,
\begin{equation}
    \mathcal{F}(\tilde\Delta) = 
        \aver{S - \tilde S}_{\tilde S} - \ln \tilde Z + \ln Z = \min,
\end{equation}
\begin{equation*}
    \tilde Z \equiv \int e^{-\tilde S}\mathcal{D}[\bar f, f], \quad
    Z \equiv \int e^{-S}\mathcal{D}[\bar f, f].
\end{equation*}

We are going to calculate the variation of Feynman's functional with respect to
$\tilde\Delta(\t)$ which parametrizes the trial action
$\tilde S = S_\mathrm{at} + \tilde S_\mathrm{hyb} + S_\mathrm{dyn}$. Doing so, we
obtain an extremum condition,
\begin{multline}
    \iint_0^\beta d\t'' d\t'''\sum_\s
    \frac{\delta \tilde G_{f,\s\s}(\t'''-\t'')}{\delta\tilde\Delta(\t-\t')}\times\\\times
    [\tilde\Delta(\t'''-\t'') - \Delta(\t'''-\t'') G_X^\textrm{at}(\t'''-\t'')] = 0.
\end{multline}

An apparent solution of this equation is $\tilde\Delta(\t) = \Delta(\t)
G^\mathrm{at}_X(\t)$.
This means that the atomic limit estimate $\aver{e^{i\theta(\t)-i\theta(\t')}} \simeq
G^\mathrm{at}_X(\t-\t')$ is indeed the best ``modulation'' function within the
proposed ansatz. The procedural change from DALA+LF to the proposed approximation 
consists of using $G^\mathrm{at}_X(\t-\t')$ instead of $Z_B$ as a prefactor of
the hybridization function. Such a change does not introduce much additional
complication, yet it allows one to achieve better results for intermediate values
of the adiabatic parameter.

\begin{figure}
    \includegraphics[scale=0.3]{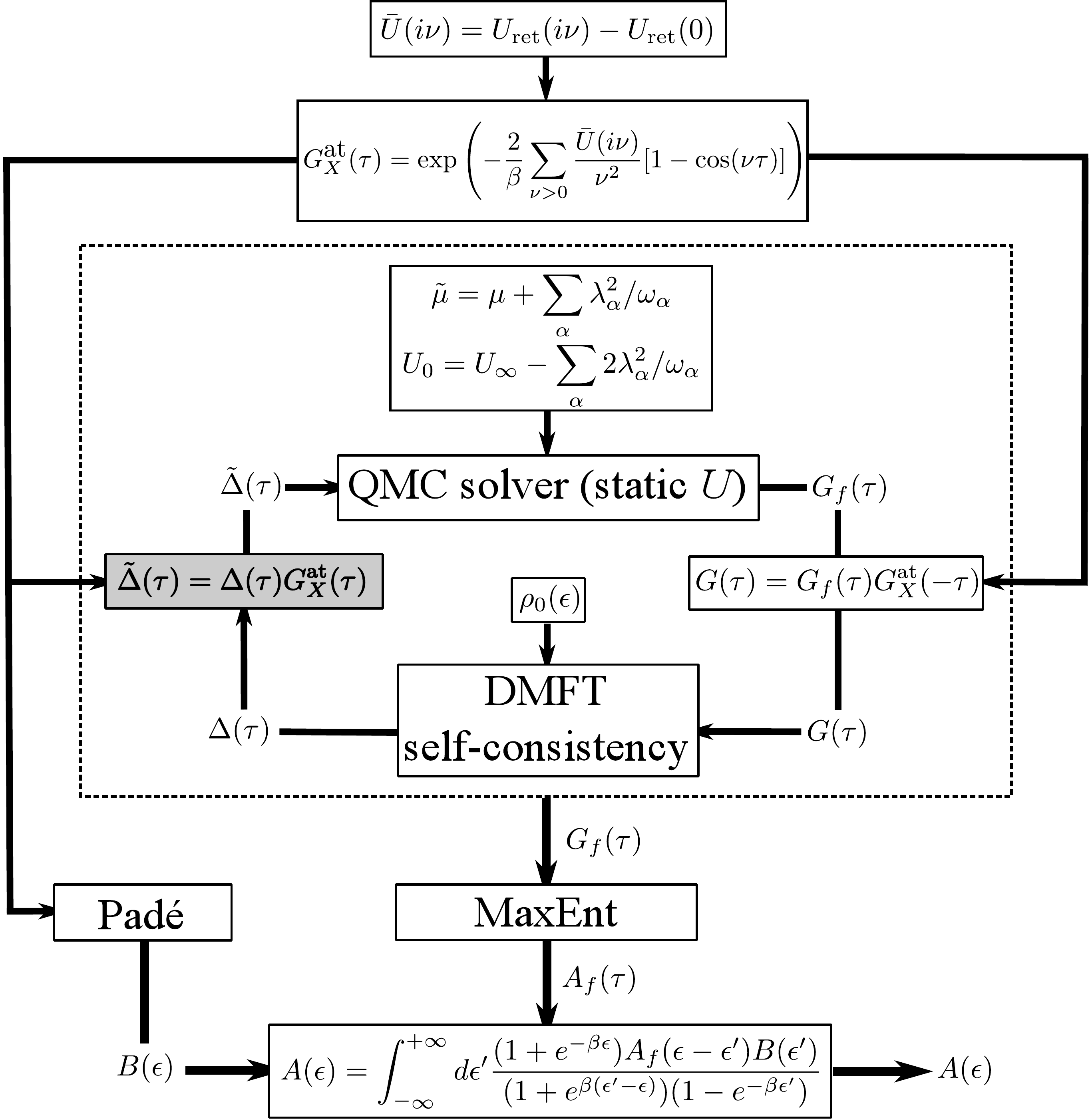}
    \caption{\label{DMFT_scheme} Scheme of the DMFT loop within the approximation
    proposed in Sec.~\ref{beyonddala}. The highlighted box would be $\tilde\Delta(\t) = \Delta(\t)$ in
    DALA and $\tilde\Delta(\t) = Z_B \Delta(\t)$ in DALA+LF.}
\end{figure}

The proposed approach is summarized in Fig. \ref{DMFT_scheme}.
The part of the scheme inside the dashed box is a DMFT loop 
involving only spinon degrees of freedom. The effective Anderson impurity model
is solved with an impurity model solver for static interactions,
e.g., a standard continuous-time quantum Monte Carlo  
(CTQMC) solver. The screened values $U_0$, $\tilde\mu$ and an
arbitrary initial guess for $\Delta$ are used as input parameters
of the first solver run. The resultant spinon Green's function $G_f(\t)$ is
then multiplied by $G_X^\textrm{at}(\t)$ in order to obtain the impurity
Green's function of physical electrons.

In the next step, a standard DMFT self-consistency procedure is performed and
an updated hybridization function 
$\Delta(\t)$ for the electrons 
is constructed. 
Within the optimized BFA approximation, $\Delta(\t)$ is replaced by
$\tilde\Delta(\t) = \Delta(\t)G_X^\textrm{at}(\t)$. In this modulated form,
the hybridization function is used as input data for the next solver run. 
The loop is repeated until $G_f(\t)$ converges with a prescribed accuracy.

Schemes for DALA and DALA+LF would differ only in the way $\tilde\Delta(\tau)$
is obtained from $\Delta(\tau)$ (multiplication by 1 and $Z_B$, respectively).

\section{Interpretation of  the ``Optimal Bose factor approach''}
\label{Bose-factor-DOS}

Before heading to the actual results, we would like to 
discuss the physical meaning of the modifications included
in the ``optimal Bose factor approach.''
To this end, we start from the DMFT self-consistency condition,
which---for a lattice with a bare dispersion 
law of electrons $\epsilon=\varepsilon(\vec{k})$---reads:
\begin{multline}
    \frac{1}{i\omega_n+\mu-\Delta(i\omega_n) - \Sigma_\textrm{AHM}(i\omega_n)} =\\=
    \frac{1}{\Omega_\textrm{BZ}}\sum_{\vec{k}}
        \frac{1}{i\omega_n+\mu-\varepsilon(\vec{k})-\Sigma_\textrm{AHM}(i\omega_n)}.
\end{multline}
Here, $\Sigma_\textrm{AHM}(i\omega_n)$ is the self-energy of 
the auxiliary Anderson-Holstein
model. Unfortunately, there is no simple form of this expression written in terms
of $\tilde\Delta$, $G_f$, and $G_X^\textrm{at}$; any such equation in the frequency domain would
inevitably contain convolutions, but not products.
However, for the particular case of the Bethe lattice with
infinite coordination number the actual equation simplifies to
\begin{equation}
    \tilde\Delta(\t) = t^2 (G_X^\textrm{at}(\tau))^2 G_f(\t),
\label{delta-tilde}
\end{equation}
i.e., the DMFT-loop for the spinons is built using a 
chargon-screened hopping parameter
$t(\tau) \equiv t G_X^\textrm{at}(\t)$.
For comparison, in DALA one has the screening coefficient 
equal to $\sqrt{G_X^\textrm{at}(\t)}$
and, in DALA+LF, to $\sqrt{Z_B G_X^\textrm{at}(\t)}$.

The proposed optimized Bose factor ansatz has 
an
important advantage over DALA+LF:
It leads to a redistribution of the spectral weight of $\Delta(\tau)$ 
rather than
to a change of the full weight. Indeed, the full weight is conserved
and is equal to a discontinuous jump of $\tilde\Delta(\tau)$ at zero time:
\begin{multline}
    \tilde\Delta(-0) - \tilde\Delta(+0) =
    \Delta(-0)G^\mathrm{at}_X(-0) - \Delta(+0)G^\mathrm{at}_X(+0) =\\= 
    \Delta(-0) - \Delta(+0).
\end{multline}

\begin{figure*}
    \includegraphics[scale=0.35]{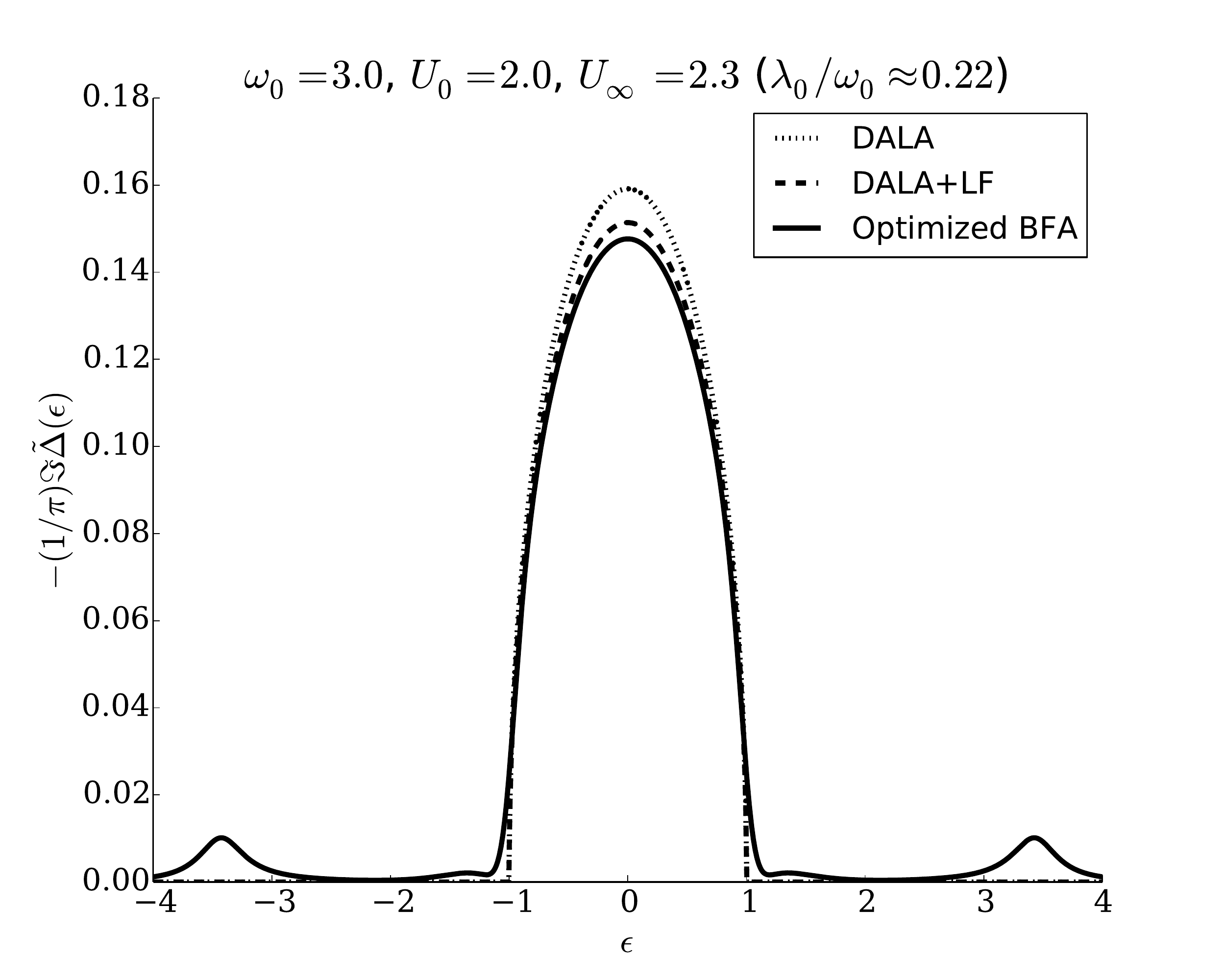}
    \includegraphics[scale=0.35]{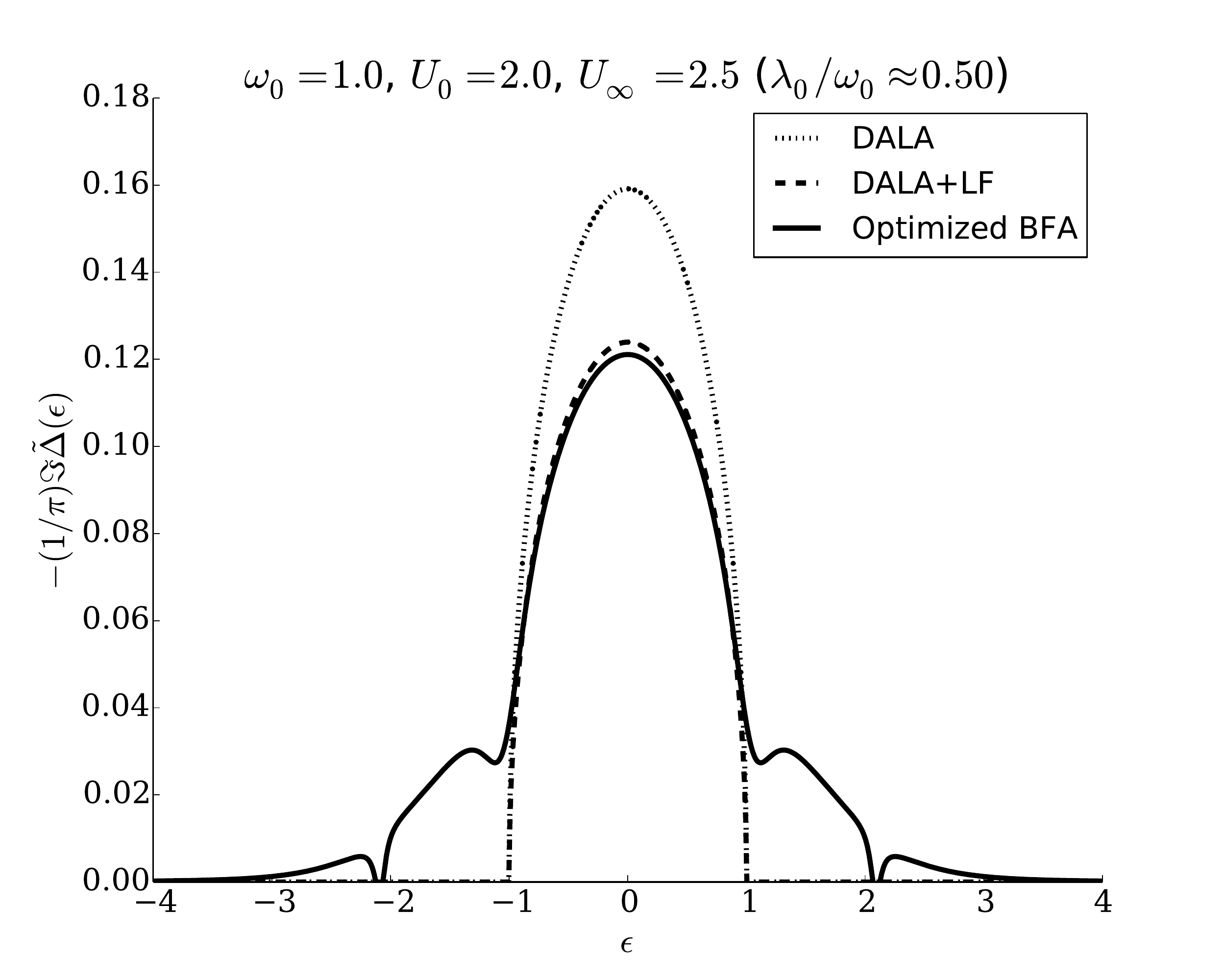}
    \includegraphics[scale=0.35]{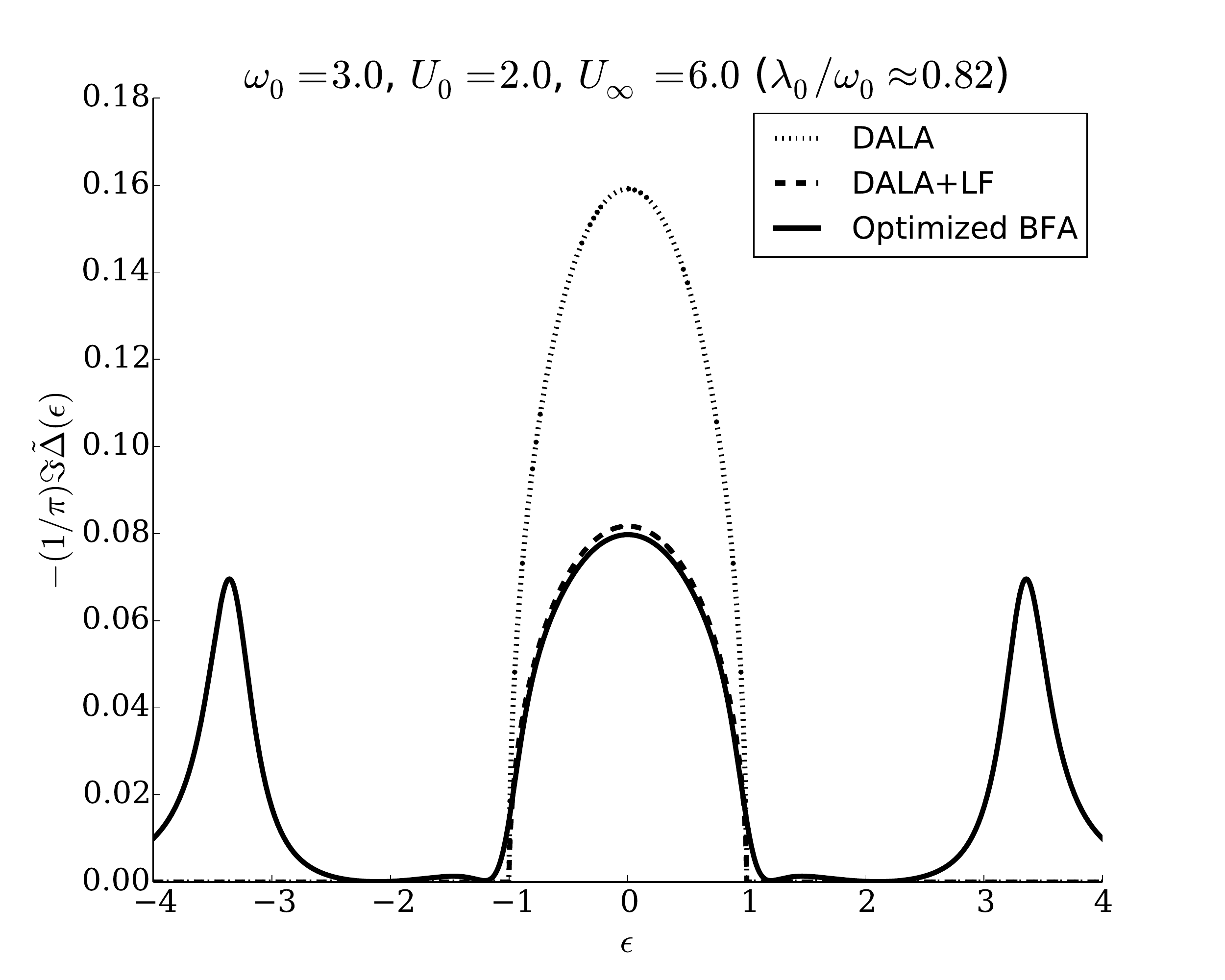}
    \includegraphics[scale=0.35]{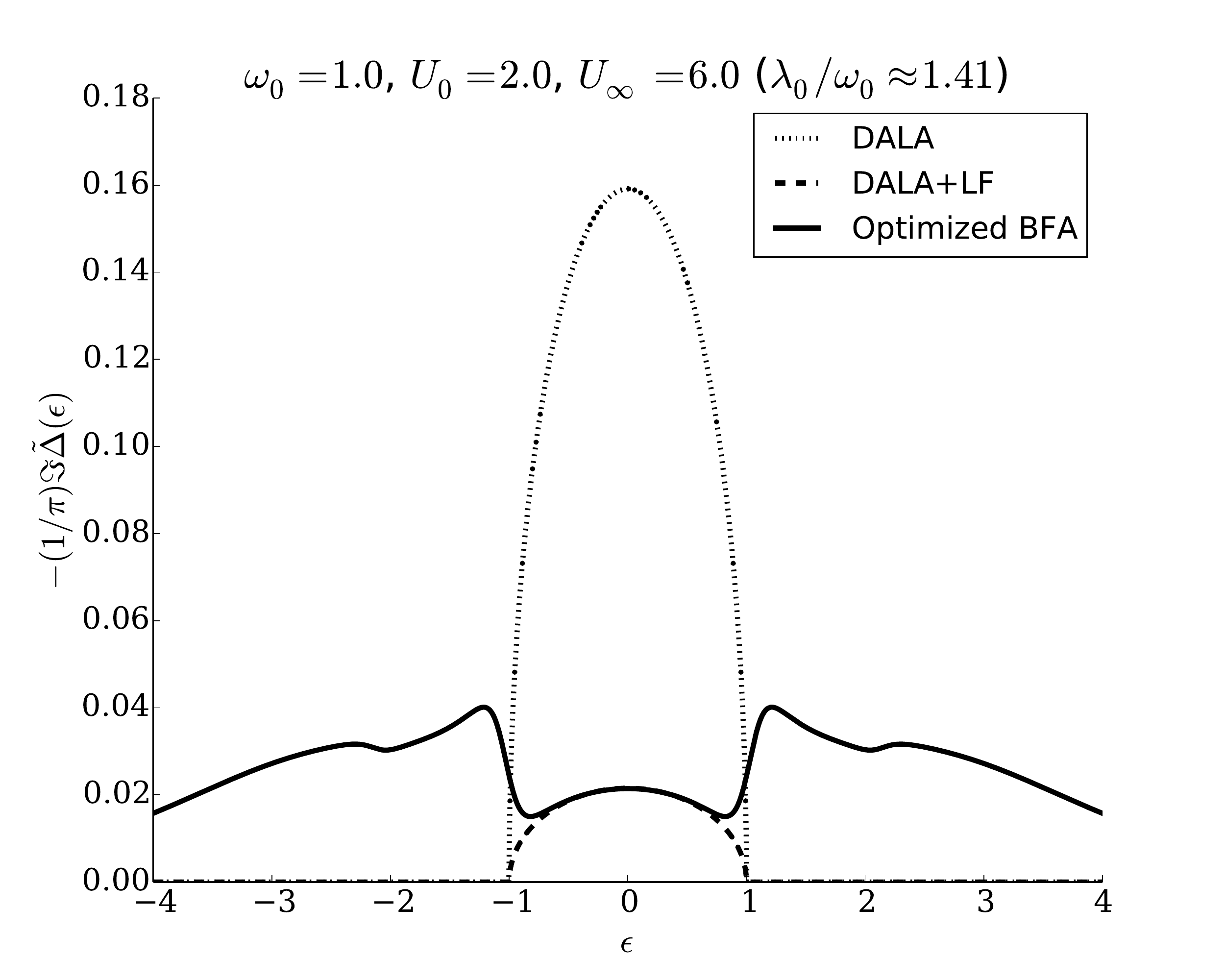}
    \caption{\label{Delta_tilde_figure} Spectral functions
    corresponding to $\tilde\Delta(\t)$ at the first iteration of the  self-consistency loop within
    different approximations and for different values
    of the adiabatic parameter $\lambda_0/\omega_0$ ($\beta=20$). The unmodulated
    (or, equivalently, DALA) spectral function is defined to have a semielliptic shape,
    $\rho(\epsilon) = \theta(1-|\epsilon|)\sqrt{1-\epsilon^2}/2\pi$.}
\end{figure*}

One can thus think of the approximation as replacing the
simplistic band renormalization by a weight-conserving transformation
of the noninteracting density of states. 
We illustrate this fact by plotting in 
Fig. \ref{Delta_tilde_figure} 
the modifications induced onto the hybridization function
corresponding to 
a model with semielliptic density of states. 
This quantity thus corresponds to the effective hybridization function
for the Bethe lattice---$\tilde\Delta(\epsilon)$ as defined in
Eq.~(\ref{delta-tilde})---at the first iteration of the above-discussed self-consistency loop.
As seen in the Figure,
there is always a low-frequency
region of the density of states (DOS) (perhaps narrow enough) in which the effect of DALA+LF 
(multiplication by $Z_B$) is almost the same as from the optimized BFA. However, the 
disagreement
grows stronger outside this region, as a value of the adiabatic parameter
$\lambda_0/\omega_0$ increases. In the deep antiadiabatic limit (upper left plot), only
a small part of the spectral weight is transferred to the plasmonic 
satellites and
DALA+LF indeed works well. For a larger value of $\lambda_0/\omega_0$ 
and a small
characteristic frequency (upper right plot), a substantial part of the 
spectral weight
is pulled out from the center of the conduction band. It is then transferred to
a newly formed pair of ``wings,'' which effectively extend the bandwidth.
The most drastic change of the DOS occurs in the 
intermediate and adiabatic regimes
(lower plots). Not only is a major part of the spectral weight transferred to
the plasmonic satellites, but also the conduction band may completely change
its shape and widen due to merging with the satellites.

\section{Technicalities}
\label{technical}

To solve the effective impurity model, we employed a hybridization expansion
Monte Carlo solver (CT-HYB), which is part of the
TRIQS application suite\cite{TRIQS}. The result of the impurity solver
is a Green's function in imaginary time, which is dressed by $G_X(-\tau)$
according to the BFA (see Fig.~\ref{DMFT_scheme}).

A note should be made about the analytic continuation procedure used.
We use Pad\'e approximants to reconstruct the bosonic spectral function $B(\epsilon)$
and Sandvik's stochastic algorithm\cite{Sandvik} for the spectrum of spinons $A_f(\epsilon)$.
Doing the convolution (\ref{convolution}) proves to be problematic due to 
the divergence
of the Bose-Einstein distribution at zero energy. 
To circumvent this difficulty, we use
an auxiliary spectral function $\tilde B(\epsilon)$, reconstructed from
$G_X^\textrm{at}(\tau) - Z_B$. By construction, $\tilde B(\epsilon)$ has a zero
at $\epsilon=0$, which compensates the divergence of the integral kernel. The resulting
density of states is immediately obtained as $A(\epsilon) = A_f \mathbf{*}\tilde B
+ Z_B A_f(\epsilon)$ (convolution in the sense of [\ref{convolution})].

\section{Comparison of different approaches}
\label{results}

\begin{figure*}
    \centering
    \begin{subfigure}[b]{0.49\textwidth}
    \includegraphics[scale=0.43]{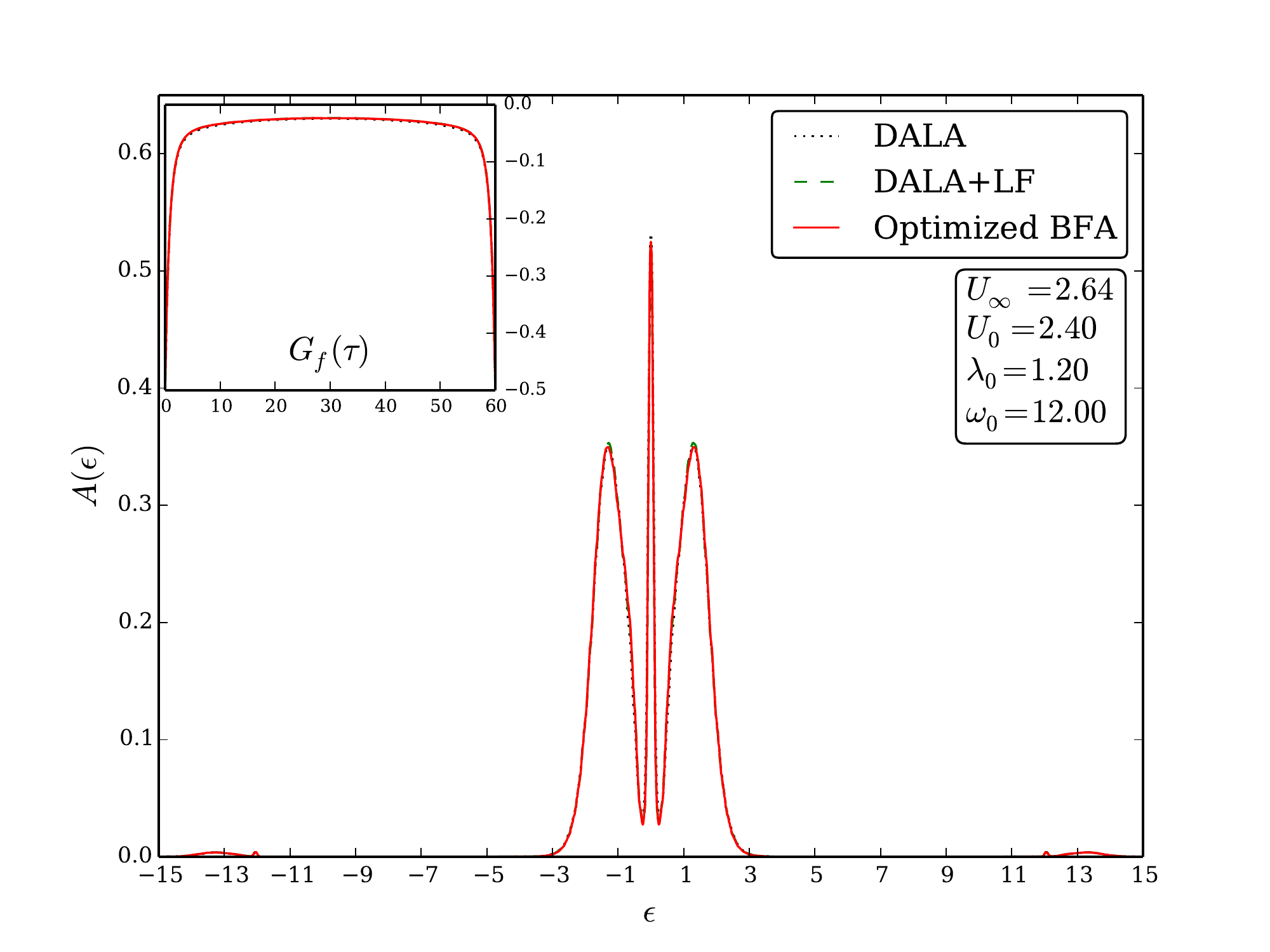}\caption{$\lambda_0/\omega_0=0.1$}
    \end{subfigure}
    \begin{subfigure}[b]{0.49\textwidth}
    \includegraphics[scale=0.43]{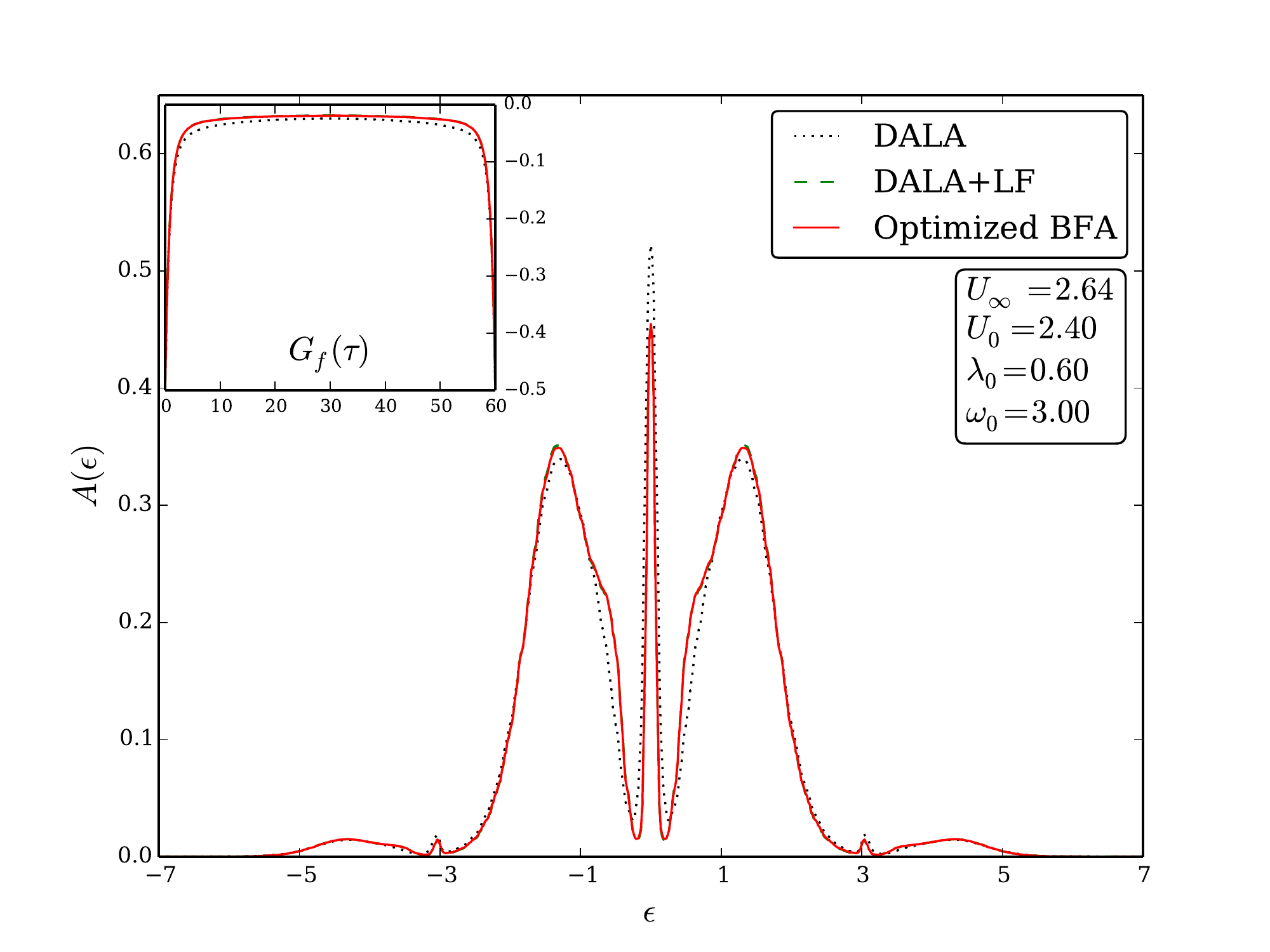}\caption{$\lambda_0/\omega_0=0.2$}
    \end{subfigure}
    \begin{subfigure}[b]{0.49\textwidth}
    \includegraphics[scale=0.43]{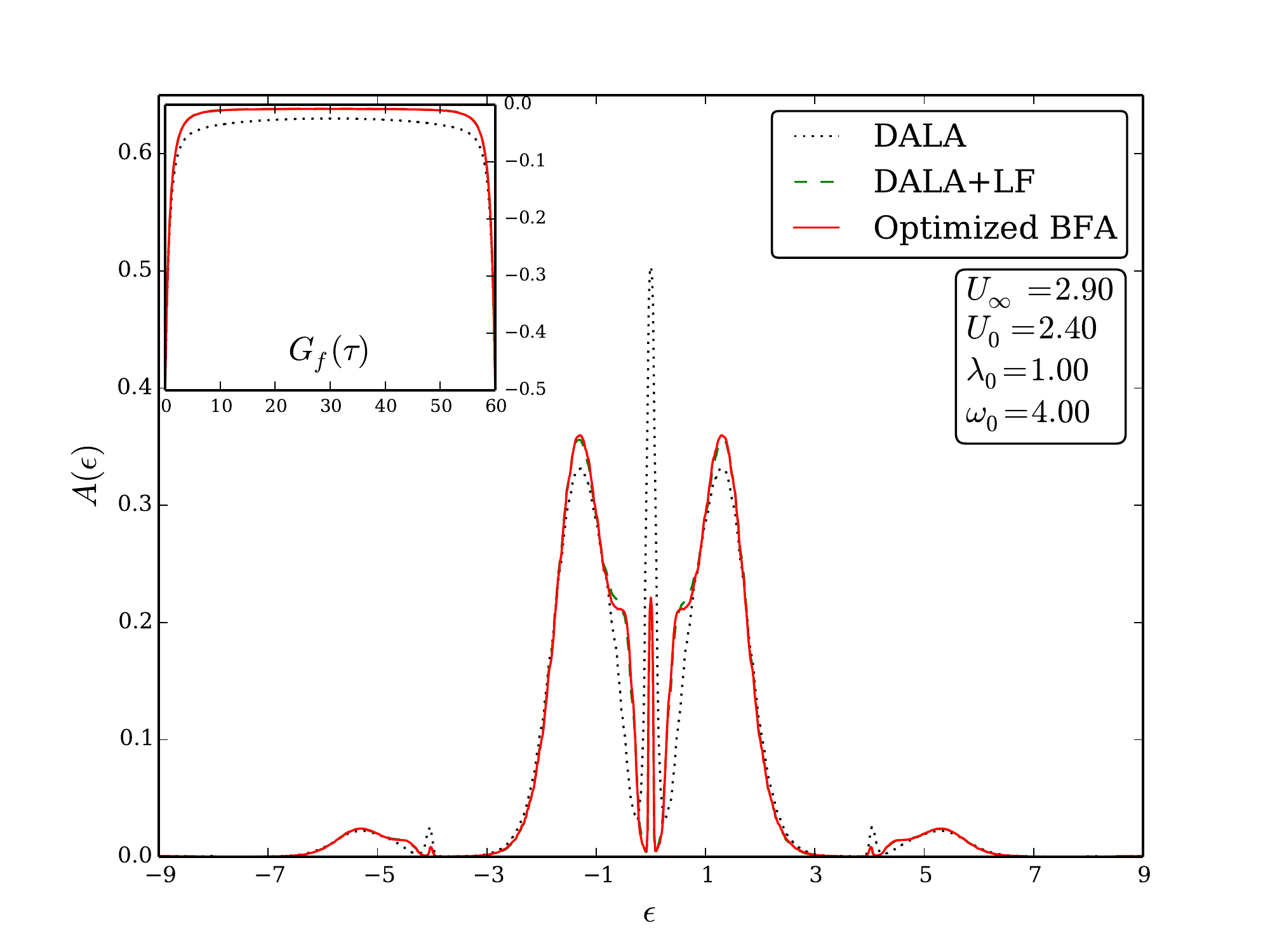}\caption{$\lambda_0/\omega_0=0.25$}
    \end{subfigure}
    \begin{subfigure}[b]{0.49\textwidth}
    \includegraphics[scale=0.43]{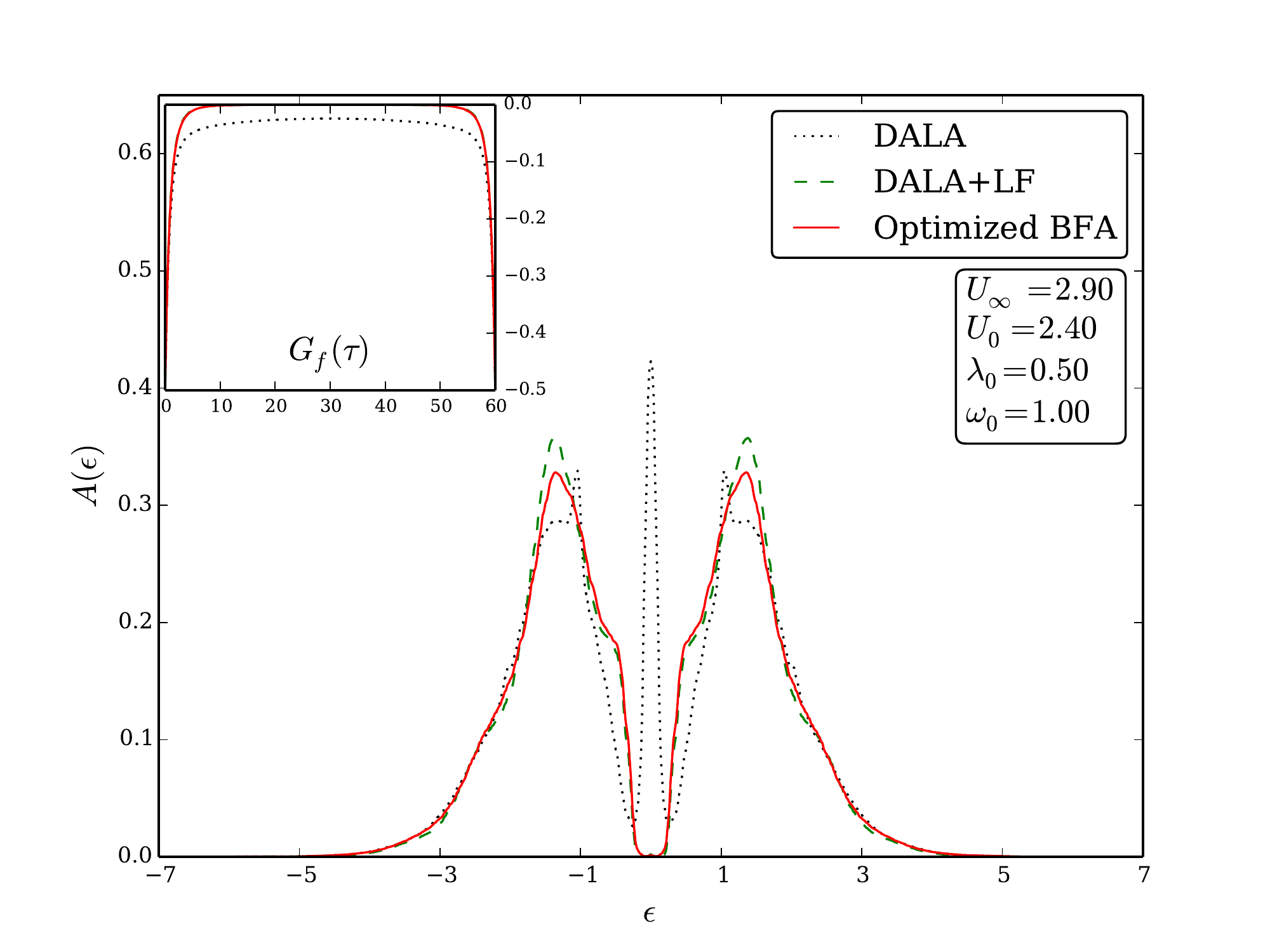}\caption{$\lambda_0/\omega_0=0.5$}
    \end{subfigure}
    \begin{subfigure}[b]{0.49\textwidth}
    \includegraphics[scale=0.43]{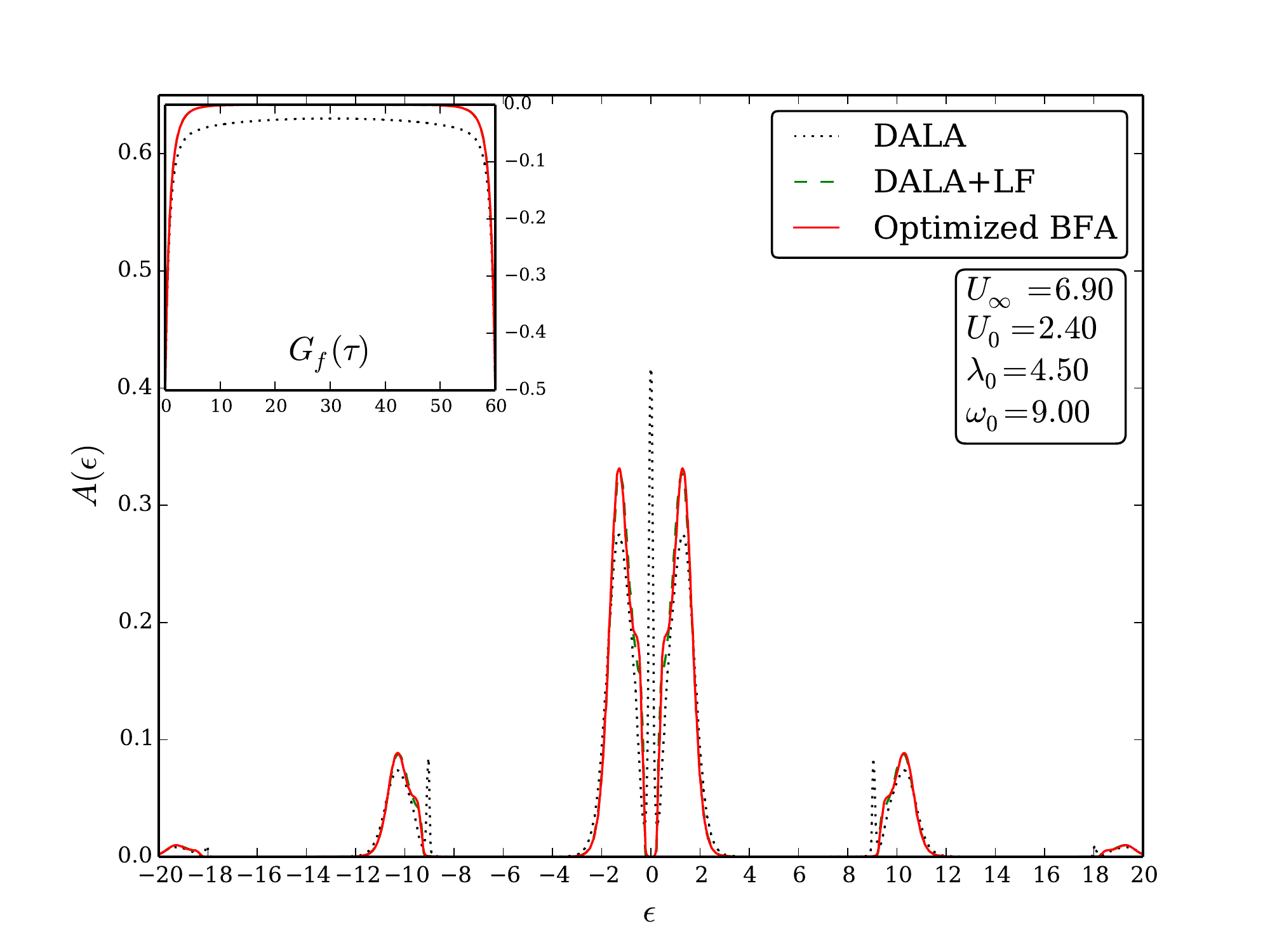}\caption{$\lambda_0/\omega_0=0.5$}
    \end{subfigure}
    \begin{subfigure}[b]{0.49\textwidth}
    \includegraphics[scale=0.43]{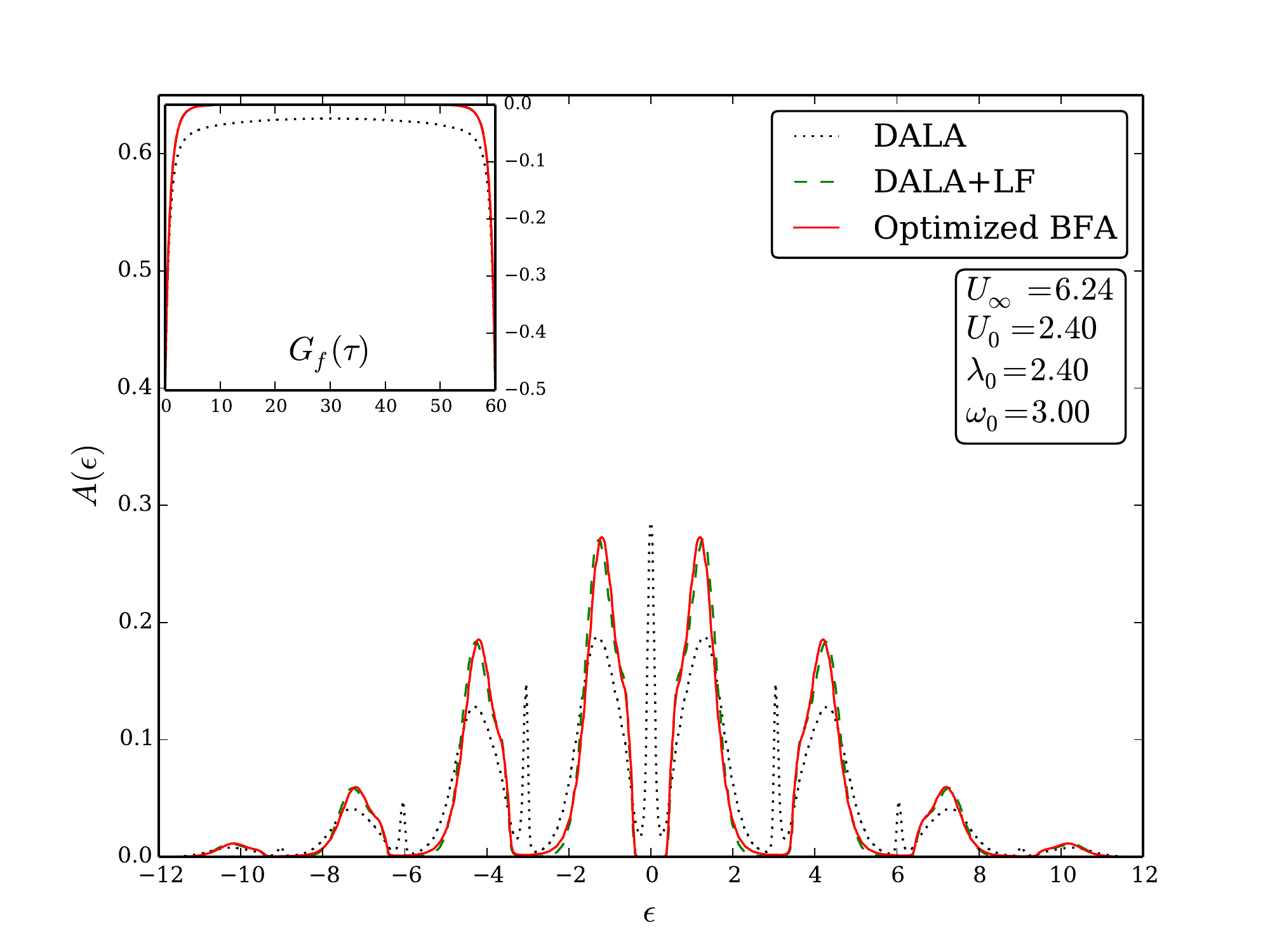}\caption{$\lambda_0/\omega_0=0.8$}
    \end{subfigure}
    \caption{\label{DOS} Local spectral functions of the Hubbard-Holstein
    model calculated within three different approximations and for six sets of parameters.
    Spinon Green's functions $G_f(\tau)$ produced by the DMFT loop are shown
    in an inset.}
\end{figure*}

In order to compare the three schemes described in the previous sections
(DALA, DALA+LF, and the optimized BFA), we have performed several series
of DMFT runs for the Hubbard-Holstein model on a Bethe lattice.

The unit of energy is set to the half bandwidth of the bare dispersion of 
the lattice.
Each of the six cases is defined by an unscreened value of on-site interaction
$U_\infty$ and by parameters of a single bosonic resonance $(\lambda_0,\omega_0)$.
Inclusion of more bosonic modes would not require a significantly larger numerical
effort. However, the resulting spectra are easier to interpret if only one bosonic
excitation is present.
The adiabatic parameter $\lambda_0/\omega_0$ is varying from 0.1 (truly antiadiabatic
regime) to 0.8 (intermediate regime).

In all cases, the value of $U_\infty$ is chosen in a way to put the
DMFT loop close to the paramagnetic Mott transition point:
$U_0=2.4$, $\beta=60$ [\onlinecite{BluemerPhD}]. 
We perform calculations at half filling, by choosing the
chemical
potential as $\mu=U_\infty/2 - 2\lambda_0^2/\omega_0$,
such as to ensure the particle-hole
symmetry of the effective Anderson model.

The spectral functions are shown in Fig. \ref{DOS}. The insets depict imaginary-time
Green's functions $G_f(\tau)$, which are directly measured by the CT-HYB solver.
Figures 4(a)-4(c) show the cases where DALA, DALA+LF, and the optimized BFA agree
on the metallic type of the solution. For a very high plasmon frequency and a small
value of the adiabatic parameter [Fig. 4(a)], the satellites are barely visible;
as the frequency approaches $U_0/2$, the satellites become more pronounced.
Spectra depicted in Figs. 4(d)-4(f) are of the insulator type as seen by DALA+LF and
by the optimized BFA, but not by DALA. Depending on the plasmon frequency, the satellites
are either completely masked by the Hubbard subbands [Fig. 4(d)] or well pronounced and
contain a considerable part of the spectral weight [Fig. 4(e)]. In the extreme case 
of Fig. 4(f) of a low frequency but strongly coupled bosons, the spectrum is comblike with the
shape of the ``teeth'' replicating the Hubbard bands.

\begin{figure}
    \begin{subfigure}[b]{0.55\textwidth}
    \includegraphics[scale=0.47]{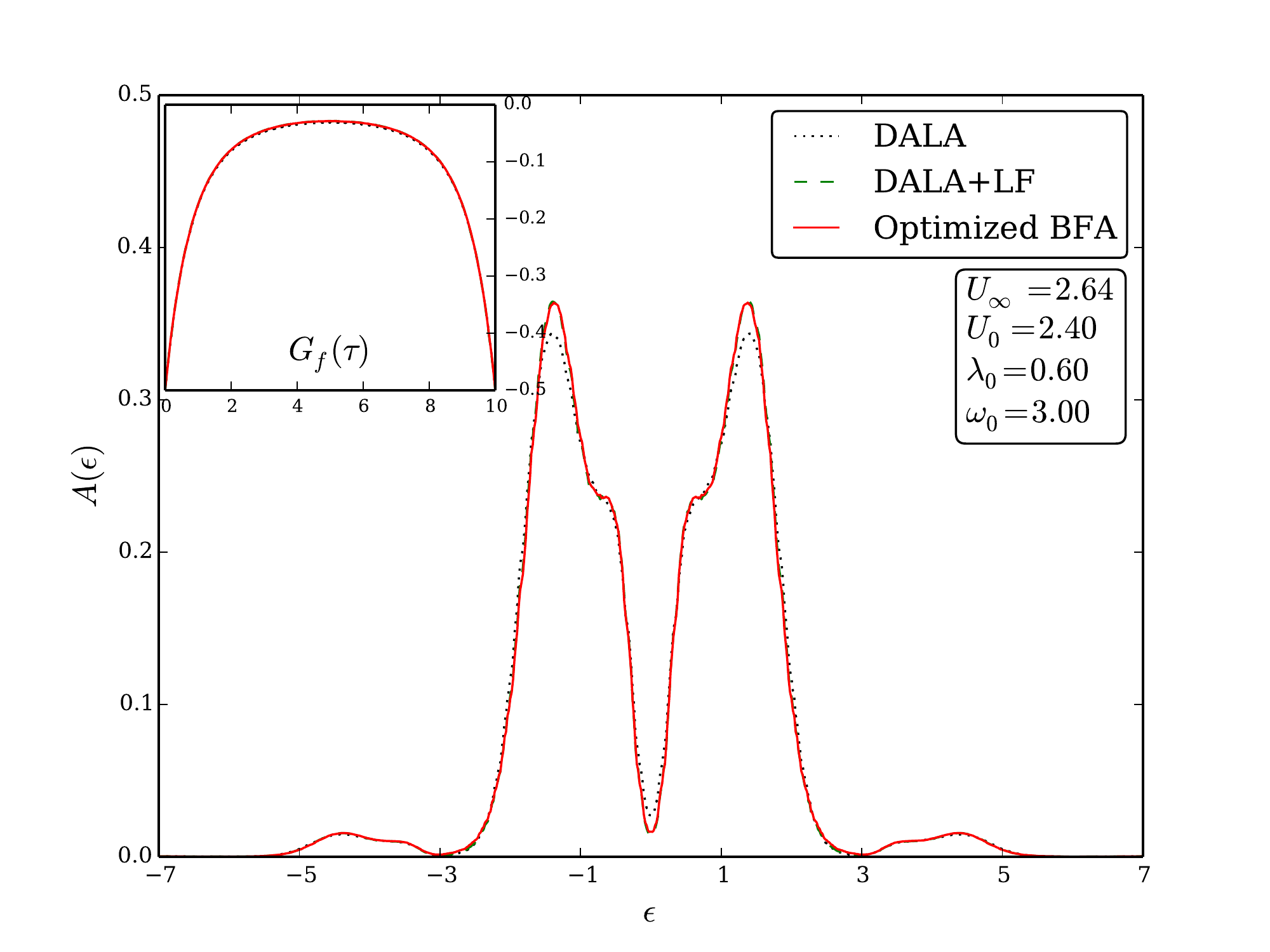}\caption{$\lambda_0/\omega_0=0.2$}
    \end{subfigure}
    \begin{subfigure}[b]{0.55\textwidth}
    \includegraphics[scale=0.47]{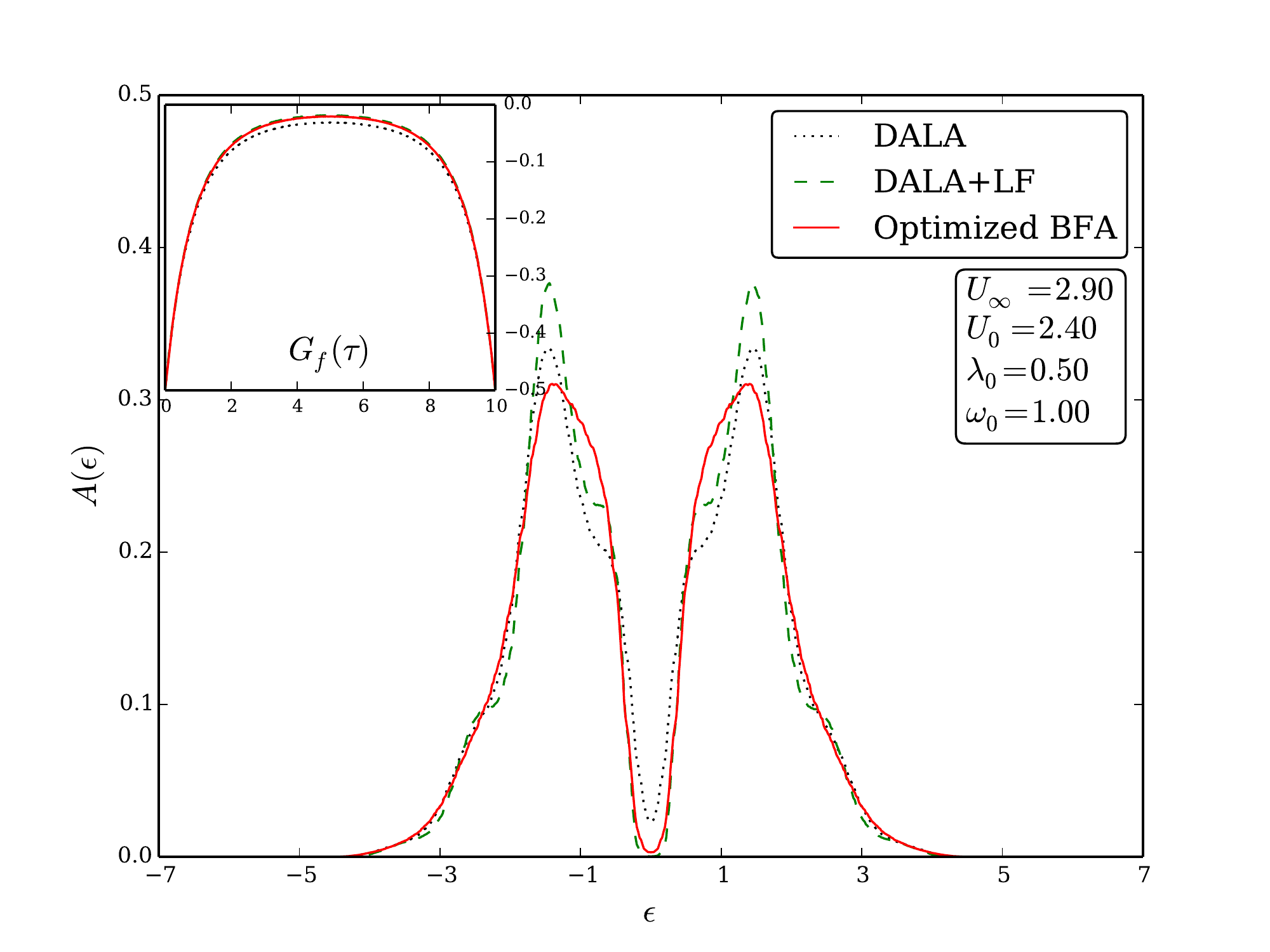}\caption{$\lambda_0/\omega_0=0.5$}
    \end{subfigure}
    \begin{subfigure}[b]{0.55\textwidth}
    \includegraphics[scale=0.47]{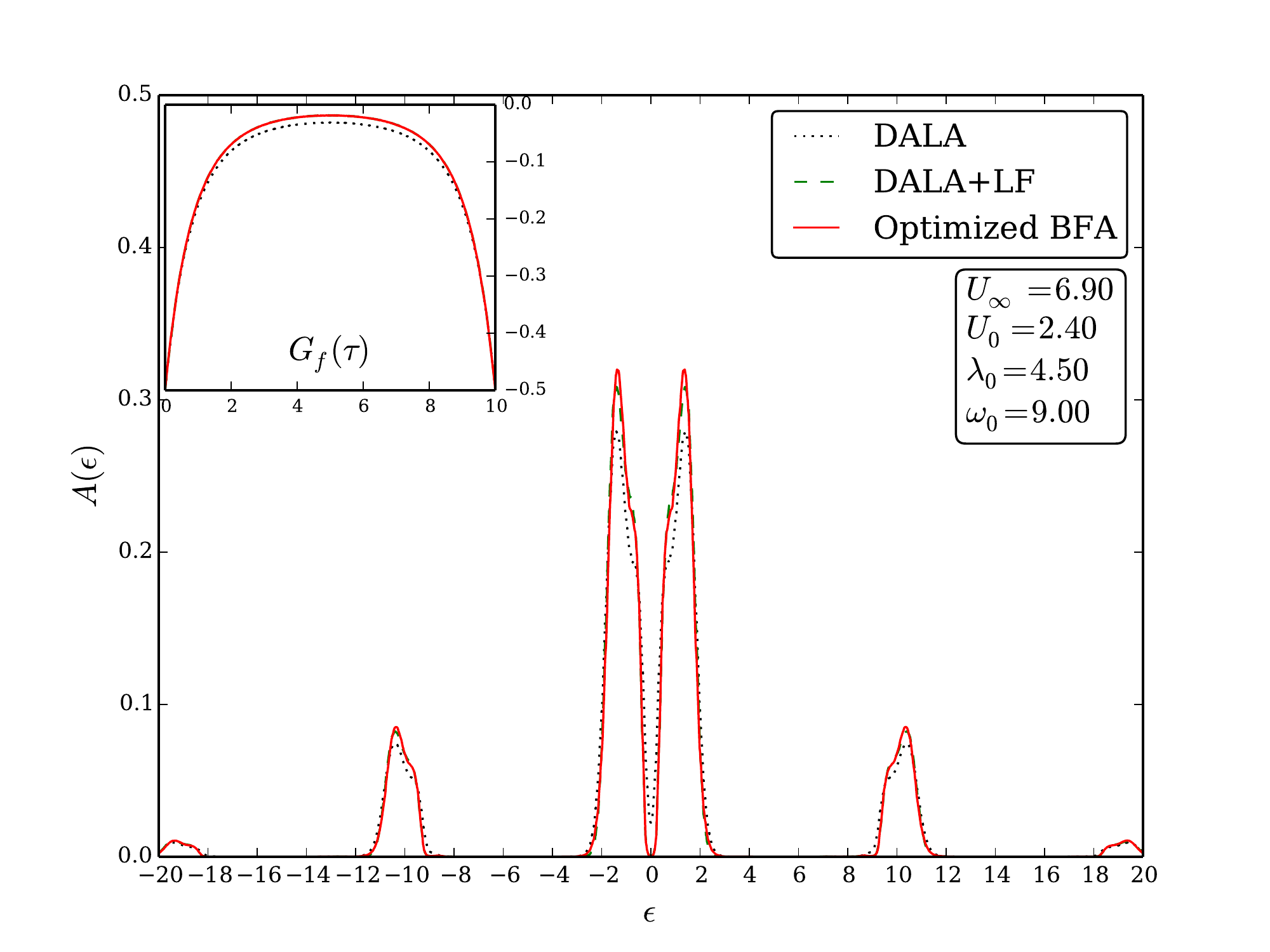}\caption{$\lambda_0/\omega_0=0.5$}
    \end{subfigure}
    \caption{\label{DOS2}
    Local spectral functions of the Hubbard-Holstein model ($\beta=10$).
    Insets: Spinon Green's functions $G_f(\tau)$ produced by the DMFT loop.}
\end{figure}

\begin{figure}
    \begin{subfigure}[b]{0.55\textwidth}
    \includegraphics[scale=0.47]{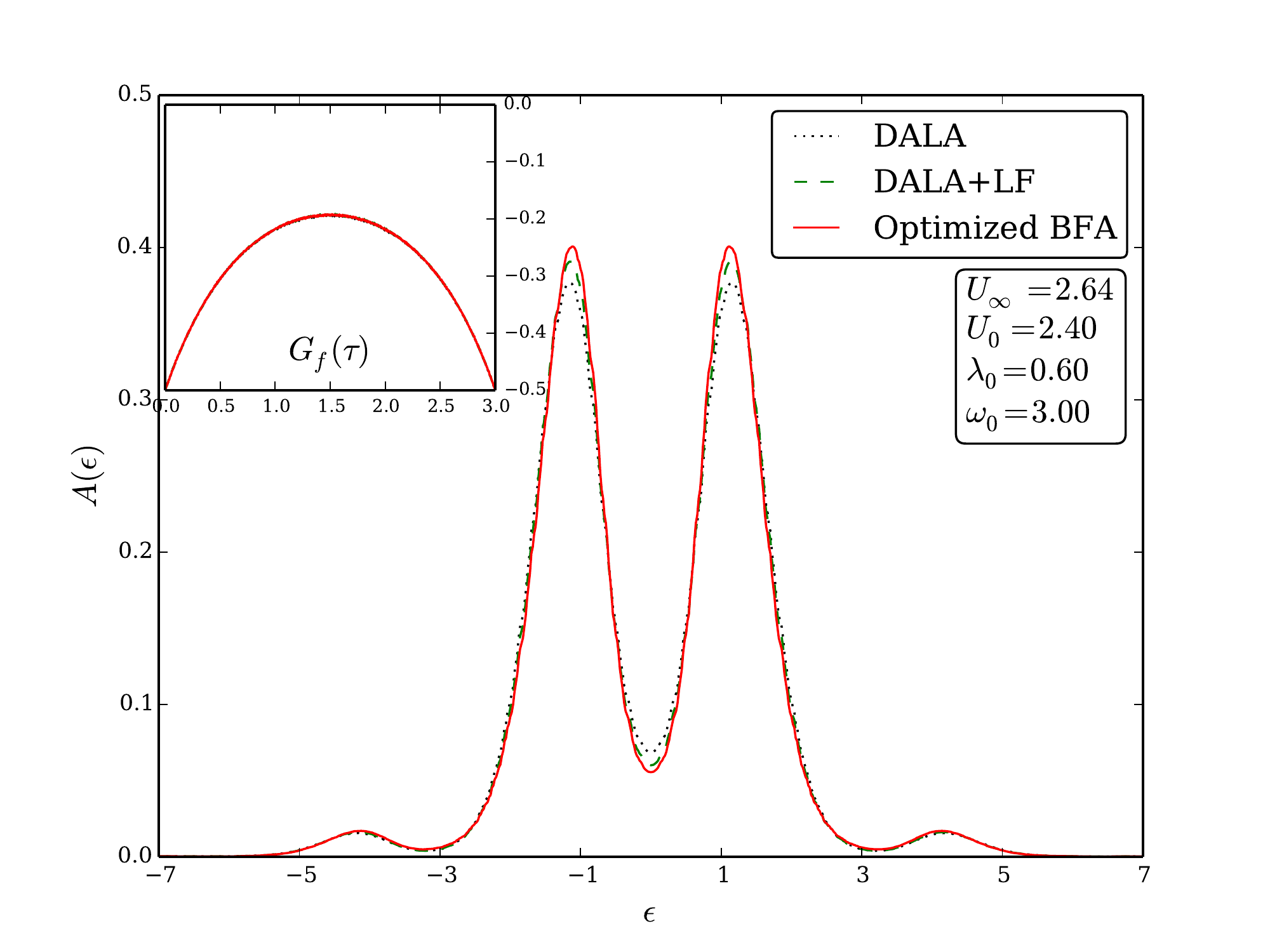}\caption{$\lambda_0/\omega_0=0.2$}
    \end{subfigure}
    \begin{subfigure}[b]{0.55\textwidth}
    \includegraphics[scale=0.47]{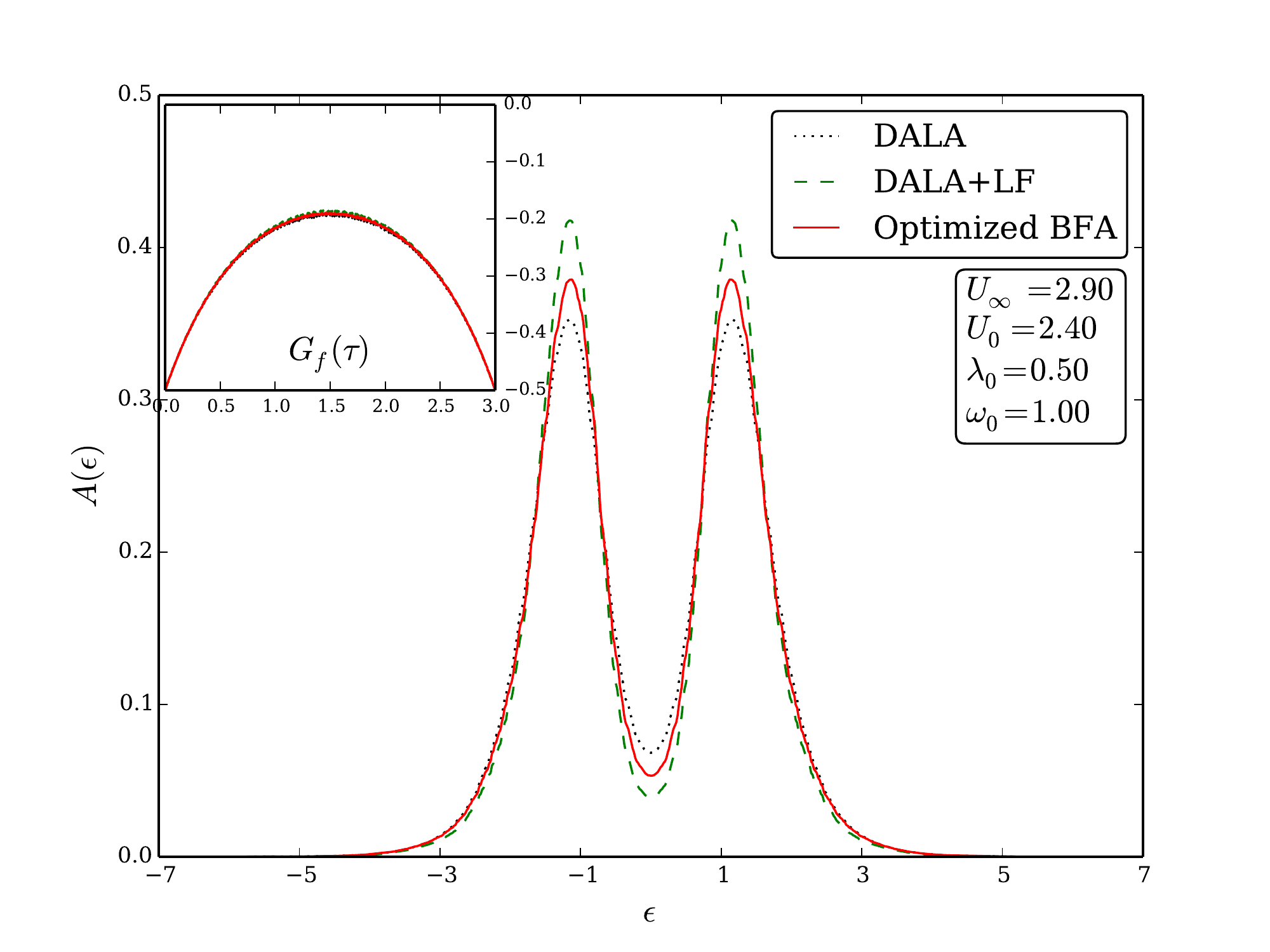}\caption{$\lambda_0/\omega_0=0.5$}
    \end{subfigure}
    \begin{subfigure}[b]{0.55\textwidth}
    \includegraphics[scale=0.47]{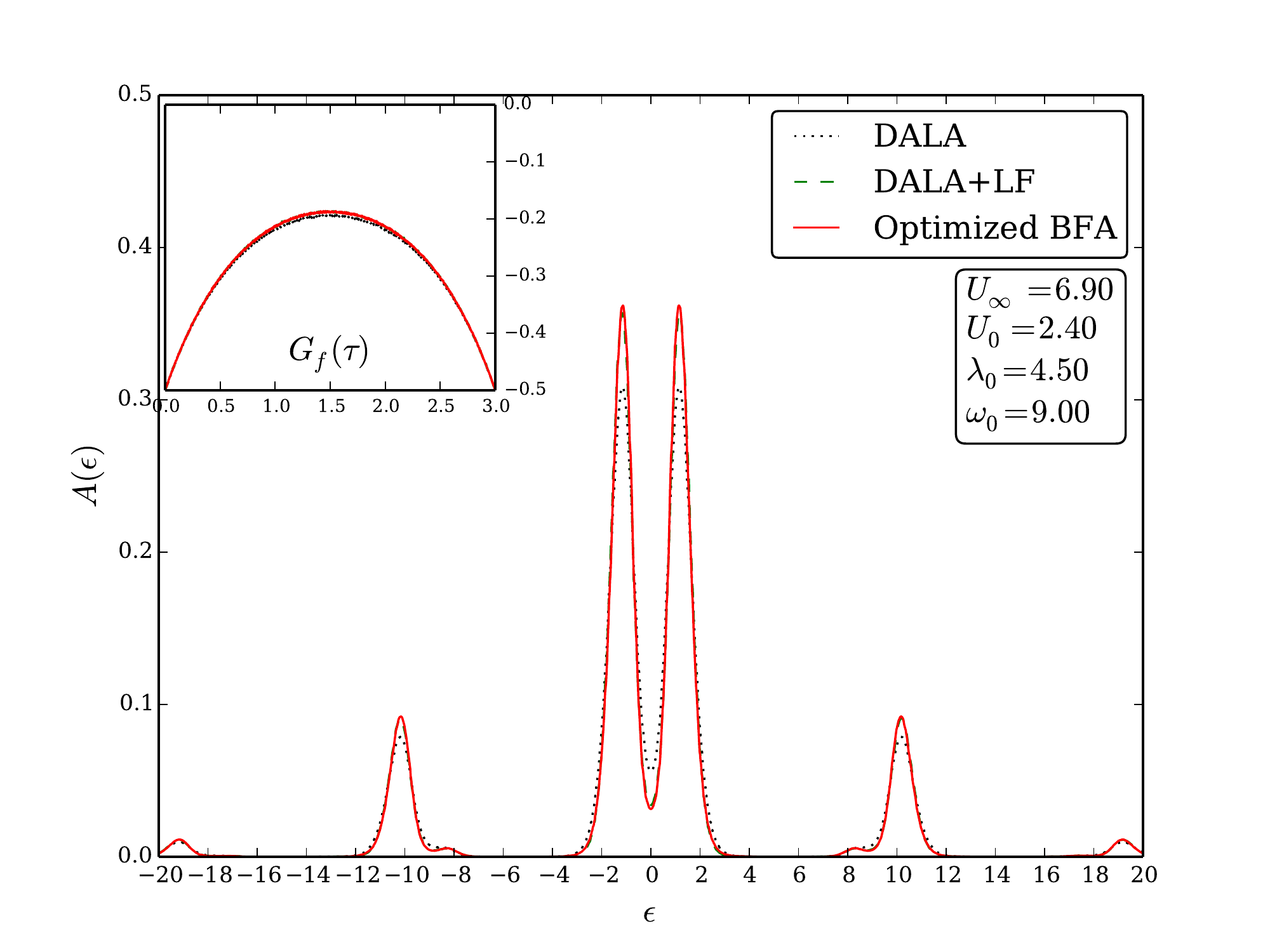}\caption{$\lambda_0/\omega_0=0.5$}
    \end{subfigure}
    \caption{\label{DOS3}
    Local spectral functions of the Hubbard-Holstein model ($\beta=3$).
    Insets: Spinon Green's functions $G_f(\tau)$ produced by the DMFT loop.}
\end{figure}

As one can see, there is a qualitative difference between DALA and the other two approximations, which
is easily understood. DALA tends to underestimate the reduction of the DOS at the Fermi level
caused by the bosons, and is thus biased towards the metallic phase.
At the same time, differences between DALA+LF and the optimized BFA are
quite subtle, if visible. The discrepancy could be more pronounced at higher temperatures,
where $G_X^\textrm{at}(\tau)$ would be approximately 
constant on a smaller part of the interval $[0;\beta]$ (see Fig. \ref{G_X_figure}). 
This issue is illustrated by spectral functions at higher temperatures presented 
in Figs. \ref{DOS2} ($\beta=10$) and \ref{DOS3} ($\beta=3$). The differences
between DALA+LF and the optimized BFA seem to be of the highest importance,
when energies of the bosonic resonance $\omega_0$ and of the atomic level $U_0/2$
are comparable. 


Finally, we discuss an interesting feature of the physical spectral functions $A(\epsilon)$
plotted in Fig. 4. Indeed, in a Fermi liquid with local self-energy---in its coherent
low-temperature regime---one expects the value of the spectral function on the Fermi
level to coincide with the value of the noninteracting density of states. This
``pinning condition'' is violated in our spectra. We checked that this is a
finite-temperature effect which is more pronounced in the optimized BFA calculations
as compared to the DALA results. The ``better'' pinning is, however, an artifact of
the underestimated correlation strength by DALA.

\section{Summary and Perspectives}
\label{conclusion}

In the present paper, we have introduced a systematic approach 
to the Hubbard-Holstein model, inspired by the slave rotor transformation 
proposed by Florens and Georges.

We have given a derivation and clarified the physical meaning of 
existing methods, such as 
the dynamic atomic limit approximation (DALA) and DALA combined with
a Lang-Firsov procedure (DALA+LF). 
DALA, being an effective tool to describe dynamic screening 
in solids, was originally
derived from an intuitively chosen ansatz. DALA+LF is an improved 
version of DALA,
which better describes the boson-induced narrowing of the 
conduction band. However,
DALA+LF suffers from a spectral weight loss issue which should be treated with care.
The proposed approach demonstrates that both DALA and DALA+LF can be understood as
simple approximations made on the fluctuating rotor-dependent factor
$\exp(i\theta(\t)-i\theta(\t'))$.

Apart from this, we have found an approximation to the exponential factor, which is
optimal in the sense of Feynman's variational criterion. This 
``optimized BFA'' approximation is closely related to DALA+LF,
but does not suffer from spectral weight loss issues. It can also be used together with
static-$U$ impurity solvers and allows one to reconstruct spectra with a 
rich resonance structure.

\begin{figure}
    \includegraphics[scale=0.3]{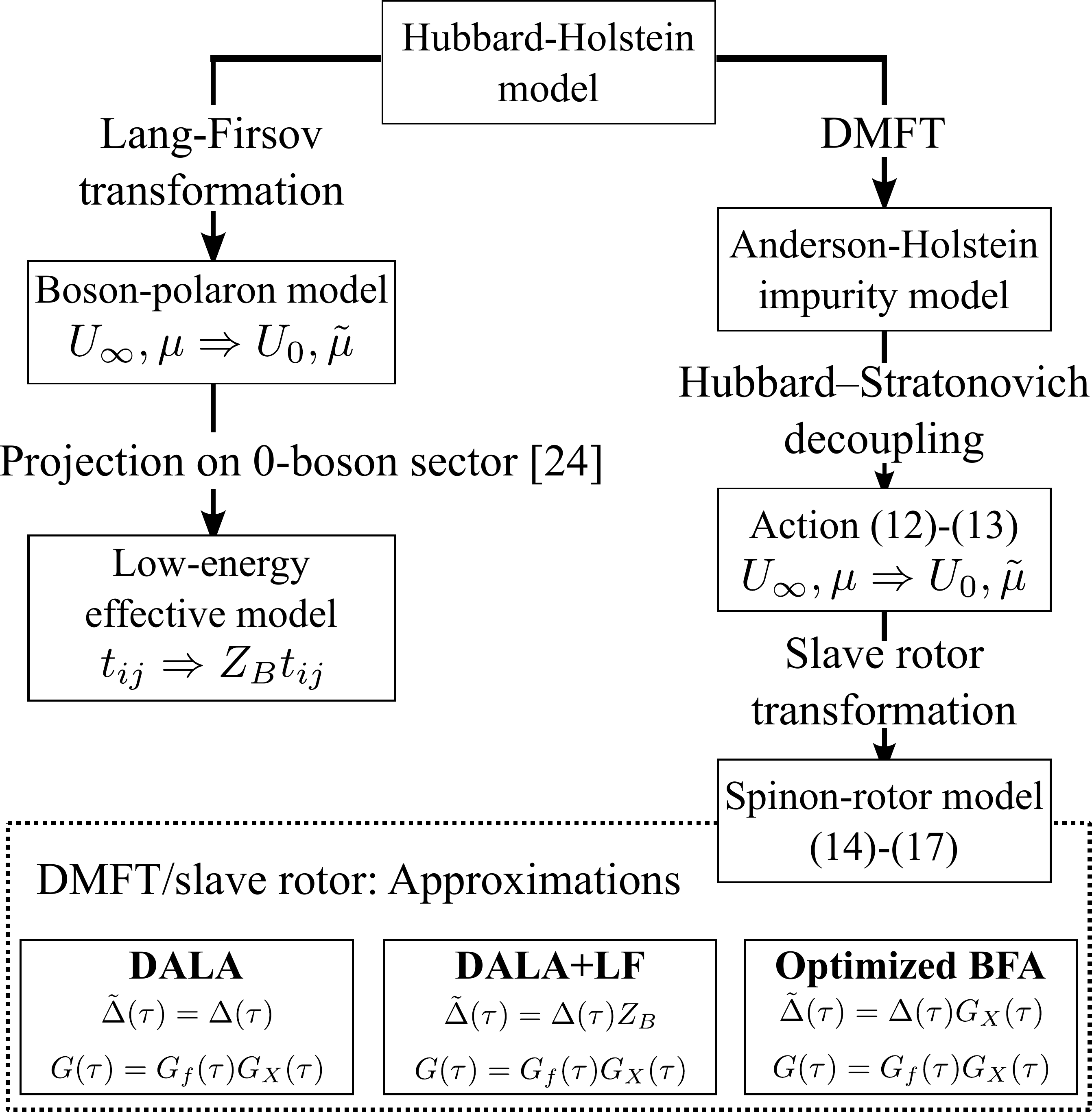}
    \caption{\label{methods} Relations between some models and methods described
    in the paper.}
\end{figure}

All main models and methods mentioned in the paper, as well as their relations, are
summarized in Fig. \ref{methods}.

A direct comparison of the methods shows that the results of the optimized BFA
are close to those of DALA+LF at low temperatures, but the difference 
should arguably
be larger for higher temperatures, when more boson excitations come into play.

An obvious further development of the proposed method will be a generalization
to multiband models. This task seems straightforward, in full analogy
to DALA. Another interesting possibility is the calculation of 
higher correlation
functions. Indeed, transformation (\ref{d_f}) together with expressions
for higher atomic correlators of rotors gives 
a simple way to calculate such quantities
[see (\ref{el_corr}) in Appendix \ref{corrfunctions}]. This could be of special
interest in the context of the dual boson approach\cite{DualBoson}.

\begin{acknowledgments}
We would like to thank A. Lichtenstein for interesting 
and inspiring discussions, and M. Casula for collaborations
around the DALA and DALA+LF schemes as implemented in [\onlinecite{DALA}].
We are grateful to Serge Florens for providing us with
his implementation of the slave rotor method.
This work was supported
by the French ANR under Projects No. SURMOTT and No. PNICTIDES, the European Research Council 
under Project No. 617196, as well as by the Deutsche Forschungsgemeinschaft
through Project No. SFB668-A3.
We acknowledge computing time by IDRIS/GENCI Orsay, under
Project No. 091393.
\end{acknowledgments}

\appendix

\section{Correlation functions in the atomic limit}
\label{corrfunctions}
A general $2n$-point correlation function of slave rotors is defined as the following
average value:
\begin{equation}
    G_X(\t_1\ldots\t_n;\t'_1\ldots\t'_n) =
    \aver{e^{i\theta(\t_1)-i\theta(\t'_1) + \ldots + i\theta(\t_n)-i\theta(\t'_n)}}.
\end{equation}
In the atomic limit, the averaging is performed with the 
action $S_\mathrm{dyn}[\theta]$
given by Eq. (\ref{S_dyn}). In spite of the fact that $S_\mathrm{dyn}$ is
quadratic in the 
phase variable $\theta(\t)$, there is no standard Wick's theorem,
since $\theta(\t)$ enters the definition of the correlation function in an
exponential form. Nonetheless, the calculation of the higher-order 
correlators is straightforward.

By definition, we have
\begin{multline}\label{int_ratio}
    G^\mathrm{at}_X(\t_1\ldots\t_n;\t'_1\ldots\t'_n) =\\=
    \frac{\int\mathcal{D}[\theta]
        e^{i\theta(\t_1)-i\theta(\t'_1) + \ldots
            + i\theta(\t_n)-i\theta(\t'_n)-S_\mathrm{dyn}[\theta]}}
    {\int\mathcal{D}[\theta] e^{-S_\mathrm{dyn}[\theta]}}.
\end{multline}

The phase field fluctuating in imaginary time is represented as a sum over
all bosonic Matsubara frequencies except $\nu=0$,
\begin{equation*}
    \theta(\t) = \frac{1}{\beta}\sum_{\nu\neq0}\theta_\nu e^{-i\nu\t}.
\end{equation*}

The ratio of the integrals in (\ref{int_ratio}) breaks up into a product over all nonzero frequencies.
Since $\theta(\t)$ is real, its Fourier components obey
the condition $\theta_\nu=\theta_{-\nu}^*$, 
and independent integration variables in the path integrals above must be
amplitudes at positive frequencies only:
\begin{equation*}
    G^\mathrm{at}_X(\t_1\ldots\t_n;\t'_1\ldots\t'_n) =
        \prod_{\nu>0} I_\nu(\t_1\ldots\t_n;\t'_1\ldots\t'_n),
\end{equation*}
\begin{multline}\label{I_def}
    I_\nu(\t_1\ldots\t_n;\t'_1\ldots\t'_n) \equiv\\\equiv
    \frac
    {\int_{-\infty}^{+\infty}d\theta'_\nu d\theta''_\nu
        e^{\frac{i}{\beta}[\theta_\nu(\sum_{k=0}^n e^{-i\nu\t_k}-e^{-i\nu\t_k'})+c.c.]
        -\frac{1}{2\beta}\frac{\nu^2|\theta_\nu|^2}{\bar U(i\nu)}}}
    {\int_{-\infty}^{+\infty}d\theta'_\nu d\theta''_\nu
        e^{-\frac{1}{2\beta}\frac{\nu^2|\theta_\nu|^2}{\bar U(i\nu)}
        }}.
\end{multline}
The integrals in (\ref{I_def}) are Gaussian and easily doable. The resulting correlation
function reads
\begin{multline}\label{G_X_result}
    G^\mathrm{at}_X(\t_1\ldots\t_n;\t'_1\ldots\t'_n) =\\=
    \exp\left[
        -\frac{1}{\beta}\sum_{\nu>0} \frac{\bar U(i\nu)}{\nu^2}
        \left|
        \sum_{k=1}^n (e^{-i\nu\t_k}-e^{-i\nu\t_k'})
        \right|^2
    \right].
\end{multline}

The transformation (\ref{d_f}) immediately gives us 
the correlator of physical electrons
through the directly measurable correlator of spinons,
\begin{multline}\label{el_corr}
    G(\t_1\ldots\t_n;\t'_1\ldots\t'_n) =\\=
    G_f(\t_1\ldots\t_n;\t'_1\ldots\t'_n)
    G^\textrm{at}_X(\t_1\ldots\t_n;\t'_1\ldots\t'_n).
\end{multline}

There is a special case for this equation, which is of high practical importance.
If all values of $\tau$ are equal to those of $\tau'$ (up to a possible index permutation),
$G_X = 1$ by definition. Therefore, averaged quantities 
such as $\aver{N(\tau)N(0)}$ or
$\aver{S_z(\tau)S_z(0)}$ are identical for physical electrons and spinons.

\section{Renormalization factor $Z_B$ at finite temperatures}
\label{Z_B_finitet}

At finite temperature, states with many bosons may effectively participate in
screening. Thus, a reasonable procedure to calculate $Z_B$
would be to average the hopping term of $\hat H_\textrm{LF}$ over all bosonic
states with a Gibbs weight distribution (for simplicity's sake, we focus on the
model with a single boson mode of energy $\omega_0$):
\begin{equation}
    Z_B t_{ij} d^\dd_{i\s} d_{j\s} \equiv
    t_{ij} \Tr[c^\dd_{i\s} c_{j\s} \hat \rho_B],
\end{equation}
\begin{equation*}\label{rho_B}
    \hat \rho_B = \frac{\exp(-\beta\omega_0\sum_i b^\dd_i b_i)}
    {\Tr[\exp(-\beta\omega_0\sum_i b^\dd_i b_i)]}.
\end{equation*}

The calculation of the trace consists of two steps. In the first step, we calculate the matrix
element of $c^\dd_{i\s} c_{j\s}$ between states with definite numbers of 
bosons $n_i$. The hopping amplitudes $t_{ij}$ are zero for $i=j$, so
the matrix element factorizes as follows:
\begin{equation}
    \melem{n_i n_j}{c^\dd_{i\s} c_{j\s} }{n_i n_j} = 
    \melem{n_i}{\hat D(\lambda/\omega_0)}{n_i}
    \melem{n_j}{\hat D(-\lambda/\omega_0)}{n_j},
\end{equation}
where $\hat D(\gamma) = \exp(\gamma b^\dd - \gamma^* b)$ is a bosonic displacement
operator. Using the Baker-Campbell-Hausdorff formula and a finite-sum representation
of the Laguerre polynomials $L_n$, we find
\begin{multline}\label{D_melem}
    \melem{n}{\hat D(-\lambda/\omega_0)}{n} =\\=
    \exp\left(-\frac{\lambda^2}{2\omega_0^2}\right)
    \sum_{k=0}^n \frac{1}{k!}
    \left(-\frac{\lambda^2}{\omega_0^2}\right)^k\binom{n}{k} =\\=
    \exp\left(-\frac{\lambda^2}{2\omega_0^2}\right)
    L_n\left(\frac{\lambda^2}{\omega_0^2}\right).
\end{multline}

In the second step of the derivation, we do an actual averaging over a thermal
state,
\begin{multline}
    Z_B =\\=
    (1-e^{-\beta\omega_0})^2 \exp\left(-\frac{\lambda^2}{\omega_0^2}\right)
    \left[
        \sum_{n=0}^\infty e^{-\beta\omega_0 n} L_n\left(\frac{\lambda^2}{\omega_0^2}\right)
    \right]^2 =\\=
    \exp\left(-\frac{\lambda^2}{\omega_0^2}\coth(\beta\omega_0/2)\right),
\end{multline}

or, in the multiple bosons case,

\begin{multline}
    \ln Z_B = -\sum_\alpha
        \frac{\lambda_\alpha^2}{\omega_\alpha^2}\coth(\beta\omega_\alpha/2) =\\=
        \frac{1}{\pi}
            \int_0^{+\infty}
            \frac{\Im U_\textrm{ret}(\epsilon)}{\epsilon^2}
            \coth(\beta\epsilon/2) d\epsilon.
\end{multline}
This is the result announced in Sec. (\ref{LF}).

\bibliographystyle{unsrtnat}

\end{document}